\documentstyle[aps,floats,psfig,twocolumn]{revtex}

\begin{document}
\tighten
\draft
%\title{Local-pair superconductivity in the one-dimensional Penson-Kolb-Hubbard
%model}
\title{${\boldmath \eta}$-superconductivity in the Hubbard chain with pair 
hopping}
\author{G.I. Japaridze $^{a,b}$, A.P. Kampf $^{a}$, M. Sekania $^{a,c}$, 
P. Kakashvili $^{c}$, and Ph. Brune $^{a}$}
\address{
$^a$ Institut f\"ur Physik, Theoretische Physik III,
Elektronische Korrelationen und Magnetismus,\\
Universit\"at Augsburg, 86135 Augsburg, Germany\\
$^b$ Institute of Physics, Georgian Academy of Sciences,
Guramishvili Str. 6, 380077, Tbilisi, Georgia\\ 
$^c$ Department of Physics, Tbilisi State University,
Chavchavadze Ave. 3, 380028, Tbilisi, Georgia 
}
\address{~
\parbox{14cm}{\rm
\medskip
The ground state phase diagram of the 1D Hubbard chain with pair-hopping 
interaction is studied. The analysis of the model is 
performed  using the continuum-limit field theory approach and exact diagonalization 
studies. At half-filling the phase 
diagram is shown to consist of two superconducting states with Cooper pair center-of-mass 
momentum $Q=0$ (BCS-$\eta_{0}$ phase) and $Q=\pi$ ($\eta_{\pi}$-phase) and four 
insulating phases corresponding to the Mott antiferromagnet, the Peierls dimerized 
phase, the charge-density-wave (CDW) insulator as well as an unconventional insulating 
phase characterized by the coexistence of a CDW and a bond-located staggered 
magnetization. Away from half-filling the phase diagram consists of the superconducting 
BCS-$\eta_{0}$ and $\eta_{\pi}$ phases and the metallic Luttinger-liquid phase. The 
BCS-$\eta_{0}$ phase exhibits smooth crossover from a weak-coupling BCS type 
to a strong coupling local-pair regime. The $\eta_{\pi}$ phase shows properties of 
the doublon (zero size Cooper pair) superconductor with Cooper pair center-of-mass momentum 
$Q=\pi$. The transition into the $\eta_{\pi}$-paired state corresponds to an abrupt change in the 
groundstate structure. After the transition the conduction band is completely 
destroyed and a new $\eta_{\pi}$-pair band corresponding to the strongly correlated doublon 
motion is created.} 
}
\maketitle
 
\section{\bf Introduction}

The problem of a crossover from a weak-coupling BCS picture of Cooper pair formation \cite{BCS} to 
a Bose-Einstein (BE) condensation of preformed local pairs has drawn special interest during the 
last two decades (see for a review Refs. \cite{MRR,Ran1}). Increased renewed 
interest to this problem mainly comes from experimental observations 
regarding the unusual properties of high-$T_{c}$ materials. Particularly important in this respect is 
the extreme short (of the order of one lattice spacing) coherence length of the pairs in the 
superconducting state \cite{Bedell} and a pseudo-gap structure in the normal-state density of states of 
underdoped cuprates \cite{PsGap}. 

Leggett \cite{Leggett} was the first to point out that BCS superconductivity can 
be continuously connected to BE condensation by increasing the two-particle attraction between
electrons from weak to strong coupling. The size of Cooper pairs shrinks continuously until 
spatially well separated bosons form, which undergo Bose condensation at a sufficiently low 
temperature. Nozieres and Schmitt-Rink \cite{Nozieres} have extended Leggett's work to lattice 
electrons and analyzed the crossover between the BCS and the BE limit as a function of coupling 
strength. The attractive ($U<0$) Hubbard model 

%%%%%%%%%%%%%%%%%%%%%%%%%%%%%%%%%%%%%%%%%%%%
\begin{equation}\label{Hubmodel}
{\cal H}= -t\sum_{n,\sigma }(c_{n,\sigma }^{\dagger }
c^{\phantom{\dagger}}_{n+1,\sigma}+ h.c.)
+ U\sum_{n} \hat{\rho}_{\uparrow}(n)\hat{\rho}_{\downarrow}(n)
\end{equation}
%%%%%%%%%%%%%%%%%%%%%%%%%%%%%%%%%%%%%%%%%%%%
was usually considered to describe the evolution from the BCS type pairing to the local pair 
(composite boson) limit \cite{RMC,Woelfle,Hous,Ran2,Santos,Sing1,Pedersen,LG,Metzner,Sing2}. 
Here $c_{n,\sigma }^{\dagger}$ ($c^{\phantom{\dagger}}_{n,\sigma }$) is the creation (annihilation) operator for an 
electron with spin $\sigma $ at site $n$ and $\hat{\rho}_{\sigma}(n)=c_{n,\sigma }^{\dagger}c^{\phantom{\dagger}}_{n,\sigma }$. 

A rather different realization of local-pairing is the $\eta$ -pairing mechanism of 
superconductivity introduced by Yang \cite{CNY1}. Yang discovered a class of eigenstates of the 
Hubbard Hamiltonian which have the property of off-diagonal long-range order (ODLRO) \cite{CNY2}, 
which in turn implies the Meissner effect and flux quantization \cite{GLS,NSZ}. 
These eigenstates are constructed in terms of doublon (on-site singlet pair) creation operators and 
can be shown to be of very high energy compared to the global ground state \cite{CNY1}. Following Yangs 
notations we call these $\eta$-paired states. On a bipartite lattice one can consider two different 
realizations of the $\eta$-paired state, constructed in terms of zero size Cooper pairs with 
{\it center-of-mass momentum} equal to zero ($\eta_{0}$-pairing) and $\pi$ ($\eta_{\pi}$-pairing), 
respectively. Yang also proved that these states cannot be ground states for the Hubbard 
model with finite interaction \cite{CNY1}. The $\eta_{0}$-superconductivity is realized in the Hubbard 
model only at infinite on-site attraction \cite{SS}. Shortly after Yang's paper, E{\ss}ler, Korepin, 
and Schoutens proposed the supersymmetric extension of the Hubbard model showing a true ODLRO and 
$\eta_{0}$-superconductivity in the ground state for a finite on-site interaction \cite{EKS1}. Later on 
several other integrable generalizations of the Hubbard model, with $\eta_{0}$ type ordering in the 
ground state were proposed \cite{EKS2}. The important common feature for all integrable models of 
$\eta$-superconductivity is the high $SU(2)\otimes SU(2)$ symmetry which is ensured by the 
{\em strongly correlated kinematics} of electrons on a lattice. 

Another model with a {\em kinematical mechanism} for the formation of Cooper pairs is the Penson-Kolb 
(PK) model \cite{PK}. The Hamiltonian of the PK model contains, in addition to the usual one-electron 
hopping term, a term that moves singlet pairs of electrons from site to site and in the one-dimensional 
case is given by
%%%%%%%%%%%%%%%%%%%%%%%%%%%%%%%%%%%
\begin{eqnarray}\label{PKmodel}
{\cal H} & = & -t\sum_{n,\alpha}(c^{\dagger}_{n,\alpha}
c^{\phantom{\dagger}}_{n+1,\alpha} + h.c.) \nonumber\\   
& + & W\sum_{n}(\hat{q}_{\uparrow }(n)\hat{q}_{\downarrow }(n)+ h.c.).
\end{eqnarray}
%%%%%%%%%%%%%%%%%%%%%%%%%%%%%%%%%%%
where $\hat{q}_{\sigma}(n)= c^{\dagger}_{n,\sigma}c^{\phantom{\dagger}}_{n+1,\sigma }$.
In the case $W<0$ the PK model (\ref{PKmodel}) describes a continuous evolution of the usual BCS type 
superconducting state at $\left|W\right| \ll t$ into a local pair $\eta_{0}$-type state at 
$\left|W\right| \gg t$ \cite{AM}. More important is that in the case of ``repulsive'' pair-hopping 
interaction ($W>0$), the $\eta_{\pi}$-paired state is realized in the ground state of the PK model for 
$W > W_{c} \simeq 2t$ \cite{BJ,BR,SA}.   

It is notable that the pair-hopping term and the Hubbard term could be obtained from the same 
general tight-binding Hamiltonian \cite{HUB} by focusing on selected terms of the two-particle 
interaction. Indeed the same matrix element of the electron-electron interaction potential $V(r)$, 
$V(n,m,k,l)=\langle n,m|V(r)|k,l\rangle$ which gives rise to the on-site Hubbard interaction for 
$n=m=k=l$ leads to the pair-hopping amplitude $W$  for $n=m$ and $k=l=n \pm 1$. Although originating 
from the same two-body potential the Hubbard and the pair-hopping couplings represent different types 
of correlations in the electron system. If the Hubbard term describes on-site correlations, the 
site-off-diagonal pair-hopping term describes part of the so-called "bond-charge" interaction. The sign 
of {\em Coulomb-driven} on-site and pair-hopping interaction is typically repulsive $U,W>0$. However, 
we will treat the parameters $U,W$ as the effective ones, assuming that they include all the possible 
renormalizations. In particular, contribution from the strong electron-phonon coupling or from the 
coupling between electrons and other electronic subsystem could give the effective attractive 
on-site ($U<0)$ interaction \cite{MRR}. The ``attractive'' ($W<0)$ pair-hopping term can originate 
from the coupling of electrons with intermolecular vibrations \cite{NWint1}, or from the on-site 
hybridization term in a generalized periodic Anderson model \cite{NWint2}.   

In this paper we investigate the interplay between these different sources for local pair formation: 
the site-diagonal Hubbard attraction and the site-off-diagonal pair-hopping interaction. We consider 
the extended Hubbard chain with pair-hopping interaction i.e. the so called Penson-Kolb-Hubbard (PKH) 
model \cite{HD}. In one-dimension the Hamiltonian reads: 
%%%%%%%%%%%%%%%%%%%%%%%%%%%%%%%%%%%%%%%%%%%%
\begin{eqnarray}\label{PKHmodel}
{\cal H}& = &-t\sum_{n,\sigma }(c_{n,\sigma }^{\dagger}c^{\phantom{\dagger}}_{n+1,\sigma}+ h.c.) + 
U\sum_{n} \hat{\rho}_{\uparrow}(n)\hat{\rho}_{\downarrow}(n) \nonumber \\
&+& W \sum_{n}(\hat{q}_{\uparrow }(n)\hat{q}_{\downarrow }(n)+h.c.) + 
\mu \sum_{n,\sigma}\hat{\rho}_{\sigma}(n). 
\end{eqnarray}
%%%%%%%%%%%%%%%%%%%%%%%%%%%%%%%%%%%%%%%%%%%%
There are $N_{e}$ particles, $N_{0}$ sites and the band filling $\nu =N_{e}/2N_{0}$ is controlled by 
the chemical potential $\mu $. In the absence of the $W$ term the Hamiltonian (\ref{PKHmodel}) 
corresponds to the Hubbard model, while in the  absence of the $U$ term it reduces to the 
Penson-Kolb model. 

The PKH model has been investigated mainly in the case $U>0$ \cite{HD,BR,Buzatu,BC,JM}. Recently 
Robaszkiewicz and Bu\l ka studied the $U<0$ PKH model by means of the Hartree-Fock approximation and 
the slave-boson mean-field method \cite{Rob}. In this paper we focus   
on the nature of phase transitions in the ground state of the PKH model between the following 
local-pair ordered phases in \cite{Rob,JM}: the $\eta_{0}$-superconductor, the 
CDW phase, and the $\eta_{\pi}$-superconductor. 

The paper is organized as follows: In Sect. II we review the model and its symmetries. We study the 
ground state phase diagram of the PKH chain using the continuum-limit bosonization approach (Sect III) 
and exact Lanczos diagonalization for chains up to $L=12$ sites (Sect. IV). Finally, Sect. V is 
devoted to a discussion and to concluding remarks. 
 
\section{\bf Review of the PKH model} 

There are two important aspects distinguishing the PKH model from the Hubbard model: the 
symmetry of these models and the nonlocal character of the pair-hopping interaction. 

Let us first consider the {\em symmetry} aspect. The PKH model and the Hubbard model are 
characterized by the same $SU(2)$-spin symmetry. The difference lies in the symmetry of the 
corresponding charge sectors - $SU(2)$ in the case of the 1/2-filled Hubbard model and 
$U(1)$ in the case of the PK model. This can be easily 
seen for a strong Hubbard attraction $|U|\gg|W|,t$ where a large gap of order $|U|$ exists 
in the single-particle excitation spectrum. Projecting the system on the subspace 
excluding single occupancy of sites and using the standard second-order perturbation theory with 
respect to $t^{2}/|U|$ \cite{Auerbach} one obtains the following effective spin-1/2 $XXZ$ spin chain 
Hamiltonian 
%%%%%%%%%%%%%%%%%%%%%%%%%%%%%%%%%%%%%%%%%%%%
\begin{eqnarray}
{\cal H}=E_{0}&+&J\sum_{n}[\frac{1}{2}(\eta_{0}^{+}(n)\eta_{0}^{-}(n+1)+h.c.)\nonumber\\
&+&\Delta \eta_{0}^{z}(n)\eta_{0}^{z}(n+1)],
\label{spinhamilt}
\end{eqnarray}
%%%%%%%%%%%%%%%%%%%%%%%%%%%%%%%%%%%%%%%%%%%%
where $E_{0}=-\frac{1}{2}N_{0}|U|$,
%%%%%%%%%%%%%%%%%%%%%%%%%%%%%%%%%%%%%%%%%%%%
\begin{equation}
J =2W -\frac{4t^{2}}{|U|}, \ \Delta = \frac{4t^{2}/|U|}{\left|{2W - 4t^{2}/|U|}\right|}
\end{equation}
%%%%%%%%%%%%%%%%%%%%%%%%%%%%%%%%%%%%%%%%%%
and the pseudospin operators are 
%%%%%%%%%%%%%%%%%%%%%%%%%%%%%%%%%%%%%%%%%%%%%
\begin{eqnarray}
\eta_{0}^{+}(n)&=& c_{n,\uparrow }^{\dagger}c_{n,\downarrow }^{\dagger}, 
\hskip 0.3cm \eta_{0}^{-}(n)=c_{n,\downarrow}c_{n,\uparrow},\nonumber\\
\\
\eta_{0}^{z}(n)&=&(c_{n,\uparrow }^{\dagger}c_{n,\uparrow}+
c_{n,\downarrow}^{\dagger}c^{\phantom{\dagger}}_{n,\downarrow }-1)/2.\nonumber
\end{eqnarray}
%%%%%%%%%%%%%%%%%%%%%%%%%%%%%%%%%%%%%%%%%%%%

As we see the charge sectors of the half-filled PKH and Hubbard models are governed by the $U(1)$ 
($\Delta \neq 1$) and the $SU(2)$ ($\Delta=1$) symmetry of the equivalent Heisenberg model 
respectively. In the case of the half-filled Hubbard model, due to the $SU(2)$-charge symmetry, the 
superconducting order 
mixes with the charge density order. The singlet superconducting (SS) and the charge density wave (CDW) 
correlations show an identical power-law decay in the infrared (large distance) limit \cite{FK}. 

For $W \neq 0$ ($\Delta \neq 1$) the spin $\eta=1/2$ antiferromagnet, with Hamiltonian 
(\ref{spinhamilt}) is in a {\em gapless planar} $XY$ phase for $\Delta \leq 1$ and in the 
{\em gapped N\'eel phase} for $\Delta > 1$ \cite{LutPesch}. The N\'eel phase is realized for 
$0<W<4t^{2}/|U|$ and corresponds to the insulating long-range-ordered (LRO) CDW state in terms of the 
initial electron system. The gapless planar $XY$ phase is realized for $W<0$ and 
$W>4t^{2}/|U|$. The tendency towards in-plane magnetic ordering reflects to the superconducting 
ordering within the initial electron system. 

In the ultimate limit $t^{2}/|U| \rightarrow 0$ the PKH model is equivalent to the spin-1/2 $XY$ model. 
Electrons only appear in singlet pairs on the same site, the interaction simply acts as a hopping 
term for these pairs. The model is $\eta$-superconducting by construction. The CDW correlations 
%%%%%%%%%%%%%%%%%%%%%%%%%%%%%%%%%%%%%%%%%%%% 
\begin{equation}\label{cdwcor}
\langle \eta_{0}^{z}(0)\eta_{0}^{z}(n)\rangle \simeq (-1)^{n}\cdot n^{-2}
\end{equation}
%%%%%%%%%%%%%%%%%%%%%%%%%%%%%%%%%%%%%%%%%%%%
decay faster then the $\eta$-superconducting correlations
%%%%%%%%%%%%%%%%%%%%%%%%%%%%%%%%%%%%%%%%%%%% 
\begin{equation}\label{sscor}
\langle \eta_{0}^{+}(0)\eta_{0}^{-}(n)\rangle \simeq  n^{-1/2}.
\end{equation}
%%%%%%%%%%%%%%%%%%%%%%%%%%%%%%%%%%%%%%%%%%%%

This picture holds true for arbitrary $W$. However, if the $\eta_{0}$-ordering is realized for $W<0$, 
then in the case of a ''repulsive'' ($W>0$) pair-hopping coupling , the $\eta_{\pi}$-superconducting
correlations 
%%%%%%%%%%%%%%%%%%%%%%%%%%%%%%%%%%%%%%%%%%%%
\begin{equation}\label{etacor}
\langle \eta_{\pi}^{+}(0)\eta_{\pi}^{-}(n)\rangle \simeq (-1)^{n}\cdot n^{-1/2}
\end{equation}
%%%%%%%%%%%%%%%%%%%%%%%%%%%%%%%%%%%%%%%%%%%%
have to dominate. In deriving (\ref{etacor}) the unitary transformation 
$\eta_{0}^{+}(n) \rightarrow (-1)^{n}\eta_{0}^{+}(n) \equiv \eta_{\pi}^{+}(n)$ has been 
used which changes the sign of the transverse exchange $J_{\perp} \rightarrow -J_{\perp}$.

Although the $XY$ model is invariant with respect to a sign change of the coupling constant, 
$W \rightarrow -W$ is not a symmetry of the PKH model. Therefore the way, the system 
approaches its limiting behaviour at $|W| \gg t^{2}/|U|$ is genuinely different for negative 
and positive $W$. 

The origin of this difference arises from the {\em site-off-diagonal nature} of the pair-hopping term. 
As far as the essence of the $W \leftrightarrow -W$ asymetry is connected to two different 
possibilities for Bose condensation of Cooper pairs with momentum $Q=0$ and $Q=\pi$ it is 
convenient to rewrite the Hamiltonian (\ref{PKHmodel}) in momentum space as 
%%%%%%%%%%%%%%%%%%%%%%%%%%%%%%%%%%%
%\begin{eqnarray}\label{MomentHam}
%{\cal H} &=& -2t\sum_{k,\sigma}c^{\dagger}_{k,\sigma}
%c^{\phantom{\dagger}}_{k,\sigma}\cos(k) \nonumber\\
%&+& \sum_{Q}(U+2W\cos(Q))A^{\dagger}_{Q}A_{Q}.
%&+& \sum_{Q}V(Q)A^{\dagger}_{Q}A_{Q}.
%\end{eqnarray}
%%%%%%%%%%%%%%%%%%%%%%%%%%%%%%%%%%%%%%%%
%%%%%%%%%%%%%%%%%%%%%%%%%%%%%%%%%%%
\begin{equation}\label{MomentHam}
{\cal H} = -2t\sum_{k,\sigma}c^{\dagger}_{k,\sigma}
c^{\phantom{\dagger}}_{k,\sigma}\cos(k) 
+ \sum_{Q}V(Q)A^{\dagger}_{Q}A^{\phantom{\dagger}}_{Q}.
\end{equation}
%%%%%%%%%%%%%%%%%%%%%%%%%%%%%%%%%%%%%%%%
Here $V(Q)=U+2W\cos(Q)$ and 
%%%%%%%%%%%%%%%%%%%%%%%%%%%%%%%%%%%
\begin{equation}\label{AqOperator}
A^{\dagger}_{Q} = {1 \over \sqrt{L}}\sum_{k}c^{\dagger}_{k,\uparrow}
c^{\dagger}_{Q-k,\downarrow}
\end{equation}
%%%%%%%%%%%%%%%%%%%%%%%%%%%%%%%%%%%%%%%%
is the creation operator for a pair of electrons with opposite spins and total 
momentum $Q$. 

In the weak-coupling (large bandwidth) limit $U,W \ll t$ only scattering processes involving 
states near the two Fermi points $k_{F} = \pm \pi\nu$ are relevant. Therefore, the scattering 
processes 
with $Q \sim 0$ (forward scattering without and with spin flip) are characterized by the effective 
coupling constant $V_{\small FS}(Q \ll k_{F}) \simeq  U+2W$ and the umklapp scattering with 
$Q \sim \pi$ by the effective amplitude coupling constant $V_{Umk}(Q \sim \pi)\simeq U-2W$. 
The umklapp scattering is irrelevant at noncommensurate band filling $\nu \neq 1/2$. Therefore, 
at $\nu \neq 1/2$ the weak-coupling phase diagram of the PKH model has to be similar to that of the 
Hubbard model with an effective interaction $U_{\em eff}=U+2W$ \cite{Buzatu,BC,JM}. At half-filling the 
umklapp scattering is relevant in the case of repulsive interaction ($U,W > 0$) and leads to the CDW 
type ordering in the PKH model at $0<U<2W$ \cite{JM}. 

In the forthcoming section we will use the 
continuum-limit bosonization approach to study the ground state phase diagram of the PKH model. 
However, already the above presented qualitative analysis indicates the absence of the 
$\eta_{\pi}$-superconducting phase in the weak-coupling phase diagram.
 
The particularity of the model in the case of strong pair-hopping interaction ($|W| \gg t$) could be 
observed even in the simple case of {\em two particles} with opposite spins on the lattice. As far as 
the total momentum of the system $Q_{tot}$ is conserved, one can treat each 
$Q$-sector of the Hilbert space independently. Since the ground state belongs to the sectors with 
$Q_{tot}=0$ or $Q_{tot}=\pi$, we restrict ourselves to these sectors only. The states with 
$Q_{tot} \neq 0$ or $\pi$ exhibit a broken time reversal symmetry and are excited states. 

Let us first consider the $U=0$ case. At $W = 0$ the ground state energy is $E_{0} = -4t$ and the 
total momentum is 
$Q_{tot} = 0$. The probability to find the on-site pair in the ground state is $1/L$. 
Eigenstates in the $Q_{tot}=\pi$ sector correspond to the highly excited states with the energy 
$E=0$. At $|W|/t \rightarrow \infty$ the $\eta_{0}$ pair given by the wave function 
$ A^{\dagger}_{0} | 0 >$ has the energy $E_{\eta_{0}}=2W$ while the $\eta_{\pi}$ pair 
given by the wave function $ A^{\dagger}_{\pi} | 0 >$ has the energy $E_{\eta_{\pi}}=-2W$. 
Therefore, for $ W < 0$, the groundstate 
{\it always} remains in the $Q_{tot}= 0$ subspace and its energy continuously varies from $-4t$ 
($W = 0$) to $-2|W|$ ($|W| \gg t$). The probability to find the on-site pair increases 
{\it continuously} 
up to 1 ($|W|/t \rightarrow \infty$). In contrast, for $ W > 0$, the energy of an $\eta_{0}$-paired 
state goes to $E_{\eta_{0}}=2W > 0$ while the energy of an $\eta_{\pi}$-paired state 
$E_{\eta_{\pi}}=-2W<0$. Thus, with increasing $W/t$, at some critical value of the pair-hopping 
coupling $W_{c}>0$, the total momentum of the system in the ground state should change from 
$Q_{tot} = 0$ to $Q_{tot} = 0$. In the $Q_{tot}= \pi$ sector the contribution of the one-particle 
hopping term to the energy vanishes. Below we will refer to this phenomenon as the 
{\em collapse of a one-particle band}. On the other hand the contribution of the delocalization 
energy for an $\eta_{\pi}$ pair is equal to $-2W$. Therefore one can roughly estimate the 
critical value to be $W_{c} \simeq -2t$. 

We now consider the effect of the Hubbard interaction. It 
is clear that the on-site repulsion $U>0$ increases the difference between the unpaired state in the 
$Q_{tot}= 0$ sector and the $\eta_{\pi}$-paired state in the $Q_{tot}= \pi$ sector. Therefore the 
critical value has to increase in the case of on-site repulsion. Contrary, the attractive Hubbard 
interaction $U<0$ supports the on-site pairing, reduces the effect of the $t$-term and therefore the 
difference between the lowest energy states in the $Q_{tot}= 0$ and $Q_{tot}= \pi$ sectors. Therefore, 
in agreement with the strong-coupling analysis the critical value has to strongly decrease in the case 
of on-site attraction. As we show below, the picture remains qualitatively similar in the case of 
many particle systems. The essence of the transition into an $\eta_{\pi}$-paired state consists of 
a collapse of the one-particle band and the creation of a strongly correlated two-particle 
$\eta$-pair band.

\section{Weak-coupling phase diagram.}

The field theory treatment of 1D systems of correlated electrons is based on
the weak-coupling approach $(|U|,\left| W\right| \ll t)$. Assuming that the
low energy physics is controlled by states near the Fermi points $\pm k_{F}$
($k_{F}a_{0}=\pi \nu $, where $a_{0}$ is the lattice spacing) we linearize
the spectrum around these points and obtain two species (for each spin
projection $\sigma $) of fermions, $R_{\sigma }(n)$ and $L_{\sigma }(n)$,
which describe excitations with dispersion relations $E(p)=\pm v_{F}p$. Here, 
$v_{F}=2ta_{0}\sin (\pi \nu)$ is the Fermi velocity and the momentum $p$ is
measured from the two Fermi points. More explicitly, one decomposes the
momentum expansion for the initial lattice operators into two parts centered
around $\pm k_{F}$ to obtain the mapping: 
%%%%%%%%%%%%%%%%%%%%%%%%%%%%%%%%%%%%%%%%%%%
\begin{equation}
c_{n,\sigma }\rightarrow {\it e}^{i\pi \nu n}R_{\sigma }(n)+e^{-i\pi \nu
n}L_{\sigma }(n),  \label{linearization}
\end{equation}
%%%%%%%%%%%%%%%%%%%%%%%%%%%%%%%%%%%%%%%%%%%
where the fields $R_{\sigma }(n)$ and $L_{\sigma }(n)$ describe right 
and left-moving particles, respectively, and are assumed to be smooth on the
scale of the lattice spacing. This allows us to introduce the continuum
fields $R_{\sigma }(x)$ and $L_{\sigma }(x)$ by 
%%%%%%%%%%%%%%%%%%%%%%%%%%%%%%%%%%%%%%%%%%%
\begin{eqnarray}
R_{\sigma }(n) &\rightarrow &\sqrt{a_{0}}R_{\sigma }(x=na_{0}),  \nonumber \\
L_{\sigma }(n) &\rightarrow &\sqrt{a_{0}}L_{\sigma }(x=na_{0}).
\label{continuumfields}
\end{eqnarray}
%%%%%%%%%%%%%%%%%%%%%%%%%%%%%%%%%%%%%%%%%%%

In terms of the continuum fields the free Hamiltonian reads: 
%%%%%%%%%%%%%%%%%%%%%%%%%%%%%%%%%%%%%%%%%%%
\begin{equation}
{\cal H}_{0}=-iv_{F}\sum_{\sigma }\int dx[:R_{\sigma }^{\dagger}\partial
_{x}R_{\sigma }:-:L_{\sigma }^{\dagger}\partial _{x}L_{\sigma }:]
\label{freelinearized}
\end{equation}
%%%%%%%%%%%%%%%%%%%%%%%%%%%%%%%%%%%%%%%%%%%
which is recognized as the Hamiltonian of a free massless Dirac field. Here and below :...: 
denotes normal ordering with respect to the ground state of the free electron 
system.

The advantage of the linearization of the spectrum is twofold: the initial
lattice problem is reformulated in terms of smooth continuum fields and --
using the bosonization procedure -- is mapped to the theory of two
independent (in the weak-coupling limit) quantum Bose fields describing
charge and spin degrees of freedom, respectively.

In terms of the continuum fields the initial lattice operators have the
form: 
%%%%%%%%%%%%%%%%%%%%%%%%%%%%%%%%%%%%%%%%%%%%%%%%%
\begin{eqnarray}
\hat{\rho}_{\sigma}(n) & \rightarrow & a_{0}
\Big\{\big(J_{R,\sigma}(x)+J_{L,\sigma}(x)\big) \nonumber \\
&+&\big(e^{-i2\pi \nu n}R_{\sigma }^{\dagger}(x)L_{\sigma }(x)+ h.c.\big)\Big\}, \label{rhoCF} \\
\hat{q}_{\sigma }(n) &\rightarrow &a_{0}\Big\{\big(e^{+i\pi \nu}J_{R,\sigma
}+e^{-i\pi \nu}J_{L,\sigma }\big)  \nonumber \\
&+&\big(e^{-i\pi \nu (2n+1)}R_{\sigma }^{\dagger}(x)L_{\sigma }(x)+ h.c.\big)\Big\},  \label{qnCF}
\end{eqnarray}
%%%%%%%%%%%%%%%%%%%%%%%%%%%%%%%%%%%%%%%%%%%%%%%%%%%
where
%%%%%%%%%%%%%%%%%%%%%%%%%%%%%%%%%%%%%%%%%%%%%%%%%%% 
\[
J_{R,\sigma }\equiv :R_{\sigma }^{\dagger}(x)R_{\sigma }(x):,\ J_{L,\sigma
}\equiv :L_{\sigma }^{\dagger}(x)L_{\sigma }(x):. 
\]
%%%%%%%%%%%%%%%%%%%%%%%%%%%%%%%%%%%%%%%%%%%%%%%%%%%
The second step is to use the standard bosonization expressions for
fermionic bilinears \cite{GNT}:
%%%%%%%%%%%%%%%%%%%%%%%%%%%%%%%%%%%%%%%%%%%%%%%%%%%
\begin{eqnarray}
-i\sum_{\sigma }\big[:R_{\sigma }^{\dagger}\partial _{x}R_{\sigma }:
&-&:L_{\sigma }^{\dagger }\partial _{x}L_{\sigma }:\big]\rightarrow  \nonumber \\
{\frac{1}{2}}\big\{(\partial _{x}\theta _{c})^{2}+(\partial _{x}\phi _{c})^{2}\big\}
&+&{\frac{1}{2}}\big\{(\partial _{x}\theta _{s})^{2}+(\partial _{x}\phi
_{s})^{2}\big\},  \label{Bos1}
\end{eqnarray}
\begin{eqnarray}
J_{R,\sigma }+J_{L,\sigma } &\rightarrow &{\frac{1}{\sqrt{2\pi }}}\big[(\partial
_{x}\phi _{c})+\sigma (\partial _{x}\phi _{s})\big],  \label{Bos2} \\
J_{R,\sigma }-J_{L,\sigma } &\rightarrow &{\frac{1}{\sqrt{2\pi }}}\big[(\partial
_{x}\theta _{c})+\sigma (\partial _{x}\theta _{s})\big]  \label{Bos3} \\
R_{\sigma }^{\dagger }(x)L_{\sigma }(x) &\rightarrow &{\frac{-i}{2\pi a_{0}}}
\exp \big({{\it i}\sqrt{2\pi }(\phi _{c}+\sigma \phi _{s})}\big).  \label{Bos4}
\end{eqnarray}
We thereby obtain 
%%%%%%%%%%%%%%%%%%%%%%%%%%%%%%%%%%%%%%%%%%%%%%%%%%%%%%%%%%%%%%%%
\begin{eqnarray}
\hat{\rho}_{\sigma }(n) &\rightarrow &a_{0}\Big\{\frac{1}{\sqrt{2\pi}}
\left[(\partial _{x}\phi _{c})+\sigma (\partial _{x}\phi _{s})\right]  \nonumber \\
&-&\frac{1}{\pi a_{0}}\sin \big({\sqrt{2\pi }(\phi _{c}+\sigma \phi _{s})-2n\pi
\nu }\big)\Big\}, \label{rhobos}\\
\hat{q}_{\sigma }(n) &\rightarrow &a_{0}\Big\{{\frac{1}{\sqrt{2\pi }}}\cos (\pi
\nu)[(\partial _{x}\phi _{c})+\sigma (\partial _{x}\phi _{s})]  \nonumber \\
&+&{\frac{i}{\sqrt{2\pi }}}\sin (\pi \nu)[(\partial _{x}\theta _{c})+\sigma
(\partial _{x}\theta _{s})]  \nonumber \\
&-&\sin \big({\sqrt{2\pi }(\phi _{c}+\sigma \phi _{s})-(2n+1)\pi \nu}\big)\Big\},
\label{qnbos}
\end{eqnarray}
%%%%%%%%%%%%%%%%%%%%%%%%%%%%%%%%%%%%%%%%%%%%%%%%%%%%%%%%%%%%%%%%
Here scalar fields $\phi _{c,s}(x)$ describe the charge and the spin degrees of freedom and 
fields $\theta _{c,s}(x)$ are their dual counterparts: $\partial _{x}\theta _{c,s}=\Pi _{c,s}$ where 
$\Pi _{c,s}$ is the momentum conjugate to the field $\phi _{c,s}$. Inserting the relations 
(\ref{rhobos})-(\ref{qnbos}) into (\ref{PKHmodel}), by virtue of the smooth variation of the fields 
$\phi_{c,s}(x)$ on the scale of lattice spacing one can convert sums into
integrals $\sum_{n}a_{0}\rightarrow \int dx$ and take the continuum limit. 

All terms in the obtained continuum-limit Hamiltonian that contain the rapidly 
oscillating phase factors $e^{\pm i2n\pi \nu}$ and $e^{\pm i4n\pi \nu}$ will drop out except at 
$\nu =1/2$, where terms containing $e^{\pm i4n\pi \nu}=1$, are not oscillating and have to be kept. 
Performing this
procedure, after rescaling of fields and lengths, the continuum-limit
version of the Hamiltonian (\ref{PKHmodel}) acquires the following form: 
%%%%%%%%%%%%%%%%%%%%%%%%%%%%%%%%%%%%%%%%%%%%%%%%%%%%%%%%%%%%%%%%
\begin{equation}
{\cal H}={\cal H}_{c}+{\cal H}_{s}  \label{bosHamiltonian}
\end{equation}
%%%%%%%%%%%%%%%%%%%%%%%%%%%%%%%%%%%%%%%%%%%%%%%%%%%%%%%%%%%%%
where 
%%%%%%%%%%%%%%%%%%%%%%%%%%%%%%%%%%%%%%%%%%%%%%%%%%%%%%%%
\begin{eqnarray}
{\cal H}_{s}&=&\int dx\Big\{{\frac{v_{s}}{2}}\left[(\partial _{x}\varphi
_{s})^{2}+(\partial _{x}\vartheta _{s})^{2}\right]\nonumber\\
&+&{\frac{m_{s}}{2\pi^{2}a_{0}^{2}}}
\cos \big(\sqrt{8\pi K_{s}}\varphi_{s}(x)\big)\Big\}  \label{SGsp}
\end{eqnarray}
%%%%%%%%%%%%%%%%%%%%%%%%%%%%%%%%%%%%%%%%%%%%%%%%%%%%%%%%%%%%%%%%
describes the spin degrees of freedom.

The charge degrees of freedom for $\nu \neq 1/2$ are described by the free
scalar field 
%%%%%%%%%%%%%%%%%%%%%%%%%%%%%%%%%%%%%%%%%%%%%%%%%%%%%%%%%%%%%%%%
\begin{equation}
{\cal H}_{c}=\frac{v_{c}}{2}\int dx\left[(\partial _{x}\varphi
_{c})^{2}+(\partial _{x}\vartheta _{c})^{2}\right],  \label{KGch}
\end{equation}
%%%%%%%%%%%%%%%%%%%%%%%%%%%%%%%%%%%%%%%%%%%%%%%%%%%%%%%%%%%%%%%%
%%%%%%%%%%%%%%%%%%%%%%%%%%%%%%%%%%%%%%%%%%%%%%%%%%%%%%%%%%%%%%%%
%\begin{equation}
%{\cal H}_{c}=\int dx\Big\{{\frac{v_{c}}{2}}[(\partial _{x}\varphi
%_{c})^{2}+(\partial _{x}\vartheta _{c})^{2}]\Big},  \label{KGch}
%\end{equation}
%%%%%%%%%%%%%%%%%%%%%%%%%%%%%%%%%%%%%%%%%%%%%%%%%%%%%%%%%%%%%%%%
and for $\nu =1/2$ by the quantum sine-Gordon field 
%%%%%%%%%%%%%%%%%%%%%%%%%%%%%%%%%%%%%%%%%%%%%%%%%%%%%%%%%%%%%%%%
\begin{eqnarray}
{\cal H}_{c}&=&\int dx\Big\{{\frac{v_{c}}{2}}\left[(\partial _{x}\varphi
_{c})^{2}+\partial _{x}\vartheta_{c})^{2}\right]\nonumber\\
&+&{\frac{m_{c}}{2\pi^{2} a _{0}^{2}
}}\cos \big(\sqrt{8\pi K_{c}}\varphi _{c}(x)\big)\Big\}.  \label{SGch}
\end{eqnarray}
%%%%%%%%%%%%%%%%%%%%%%%%%%%%%%%%%%%%%%%%%%%%%%%%%%%%%%%%%%%%%%%%
Here we have defined
%%%%%%%%%%%%%%%%%%%%%%%%%%%%%%%%%%%%%%%%%%%%%%%%%%%%%%%%%%%%%%%%
%\begin{eqnarray}
%K_{c} &\simeq& 1+{\frac{1}{2}}g_{c} \equiv  1- \frac{1}{2}\frac{U+2W}{2\pi t},  
%\label{Kc} \\
%m_{c}&=&-(U-2W)/2\pi t\label{Mc}\\
%K_{s} &\simeq& 1+\frac{1}{2}g_{s} \equiv  1+ \frac{1}{2}(U+2W)/2\pi t,
%\label{Ks} \\
%m_{s} &=& (U+2W)/2\pi t,  
%\label{Ms} \\
%v_{c} &=&v_{F}K^{-1}_{c},\hskip0.3cm v_{s}= v_{F}K^{-1}_{s}.  \label{vcvs}
%\end{eqnarray}
%%%%%%%%%%%%%%%%%%%%%%%%%%%%%%%%%%%%%%%%%%%%%%%%%%%%%%%%%%%%%%%%
%%%%%%%%%%%%%%%%%%%%%%%%%%%%%%%%%%%%%%%%%%%%%%%%%%%%%%%%%%%%%%%%
\begin{eqnarray}
K_{c} &\simeq& 1+{\frac{1}{2}}g_{c} \equiv  1- \frac{1}{2}{U+2W \over 2\pi t},  
\label{Kc} \\
m_{c}&=&-{U-2W \over 2\pi t}\label{Mc}\\
K_{s} &\simeq& 1+\frac{1}{2}g_{s} \equiv  1+\frac{1}{2}{U+2W \over 2\pi t},
\label{Ks} \\
m_{s} &=& {U+2W \over 2\pi t},  
\label{Ms} \\
v_{c} &=&v_{F}K^{-1}_{c},\hskip0.3cm v_{s}= v_{F}K^{-1}_{s}.  \label{vcvs}
\end{eqnarray}
%%%%%%%%%%%%%%%%%%%%%%%%%%%%%%%%%%%%%%%%%%%%%%%%%%%%%%%%%%%%%%%%
The mapping of the Hamiltonian (\ref{PKHmodel}) 
into the quantum theory of two independent charge and spin Bose fields, 
allows to study the ground state phase diagram of the initial electron  
system, using the far-infrared properties of the bosonic Hamiltonians (\ref{SGsp})-(\ref{SGch}).

\subsection{Renormalization group analysis}

Let us first consider the half-filled band case when both the spin
and the charge sectors of the system are governed by the quantum sine-Gordon
(SG) fields (\ref{SGsp}) and (\ref{SGch}). The infrared behavior of the SG
Hamiltonian ${\cal H}_{c,s}$ is described by the corresponding pair of
renormalization group (RG) equations for the effective coupling constants 
$M_{c(s)}$ and $K_{c(s)}$ \cite{WIE}
%%%%%%%%%%%%%%%%%%%%%%%%%%%%%%%%%%%%%%%%%%%%%%%%%%%%%%%%%%%%%%%%
\begin{eqnarray} \label{RGeq1}
{dM_{c(s)}(L) \over dL}&=&-2\big(K_{c(s)}(L)-1\big)M_{c(s)}(L)\nonumber\\ 
{dK_{c(s)}(L) \over dL}&=&-\frac{1}{2}M_{c(s)}^{2}(L),
\end{eqnarray}
%%%%%%%%%%%%%%%%%%%%%%%%%%%%%%%%%%%%%%%%%%%%%%%%%%%%%%%%%%%%%%%%
%%%%%%%%%%%%%%%%%%%%%%%%%%%%%%%%%%%%%%%%%%%%%%%%%%%%%%%%%%%%%%%%
where $L=ln(a_{0})$, $K_{c(s)}(0)=1+\frac{1}{2}g_{c(s)}$ and 
$M_{c(s)}(0)=m_{c(s)}$. Each pair of RG equations (\ref{RGeq1}) 
describes the Kosterlitz-Thouless 
transition \cite{KT} in the charge and spin channels. The flow lines lie on the 
hyperbola
%%%%%%%%%%%%%%%%%%%%%%%%%%%%%%%%%%%%%%%%%%%%%%%%%%%%%%%%%%%%%%%%
\begin{equation}
4(K_{c(s)}-1)^{2}-M_{c(s)}^{2}= \mu_{c(s)}^{2}=g_{c(s)}^{2}-m_{c(s)}^{2},  
\label{flowlines}
\end{equation}
%%%%%%%%%%%%%%%%%%%%%%%%%%%%%%%%%%%%%%%%%%%%%%%%%%%%%%%%%%%%%%%%
and -- depending on the relation between the bare coupling constants 
$g_{c(s)} \equiv 2(K_{c(s)}-1)$ and $m_{c(s)}$ -- exhibit two different regimes 
(see Fig.\ref{fig:GSFD}):

%%%%%%%%%%%%%%%%%%%%%%%%%%%%%%%%%%%%%%%%%%%%%%%%%%%%%
\begin{figure}
\vspace{0mm}
\centerline{\psfig{file=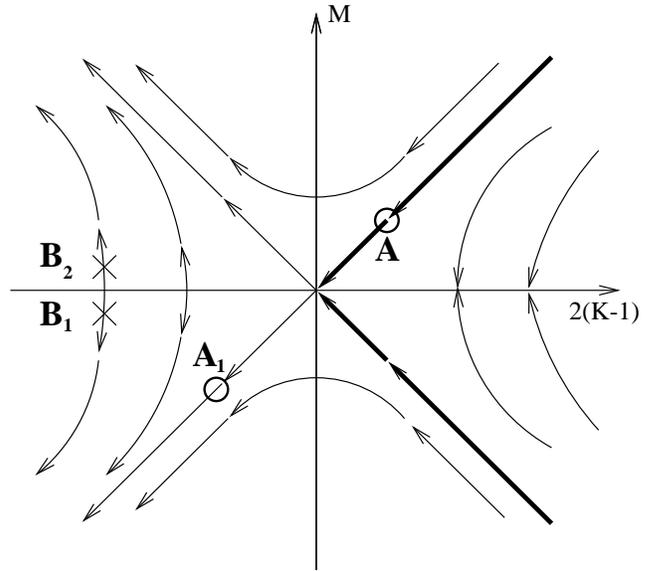,width=85mm,
silent=}}
\vspace{5mm}
\caption{Renormalization-group flow diagram; the arrows denote the 
direction of flow with increasing length scale. The circles correspond 
to the starting points in the case of a gapless (A) and a gapped ($A_{1}$) spin channel. 
The crosses correspond to the starting points for two different gapped sectors in the 
charge channel with $\langle \varphi _{c}\rangle =0$ ($B_{1}$) and 
$\langle \varphi _{c}\rangle = \pi/8K_{c}$ ($B_{2}$).}
\label{fig:GSFD}
\end{figure}
%%%%%%%%%%%%%%%%%%%%%%%%%%%%%%%%%%%%%%%%%%%%%%%%%%

For $g_{c(s)} \geq \left|m_{c(s)} \right|$ we are in the weak
coupling regime; the effective mass $M_{c(s)} \rightarrow 0$. The low energy
(large distance) behavior of the gapless charge (spin) excitations is
described by a free scalar field. The corresponding correlations show a
power law decay 
%%%%%%%%%%%%%%%%%%%%%%%%%%%%%%%%%%%%%%%%%%%%%%%%%%%%%%%%%%%%%%%%
\begin{eqnarray}
\langle e^{i\sqrt{2\pi K^{\ast }}\varphi (x)}e^{-i\sqrt{2\pi K^{\ast }}
\varphi (x^{\prime })}\rangle &\sim &\left| x-x^{\prime }\right| ^{-K^{\ast
}},  \label{freecorrelations1} \\
\langle e^{i\sqrt{2\pi /K^{\ast }}\theta (x)}e^{-i\sqrt{2\pi /K^{\ast }}
\theta (x^{\prime })}\rangle &\sim &\left| x-x^{\prime }\right| ^{-1/K^{\ast
}},  \label{freecorrelations2}
\end{eqnarray}
%%%%%%%%%%%%%%%%%%%%%%%%%%%%%%%%%%%%%%%%%%%%%%%%%%%%%%%%%%%%%%%%
and the only parameter controlling the infrared behavior in the gapless
regime is the fixed-point value of the effective coupling constants $K_{c(s)}^{\ast}$.

For $g_{c(s)}< \left|m_{c(s)} \right| $ the
system scales into a strong coupling regime: depending on the sign of the
bare mass $m_{c(s)}$, the effective mass $M_{c(s)}\rightarrow \pm \infty $,
which signals the crossover into a strong coupling regime and indicates the
dynamical generation of a commensurability gap in the excitation spectrum. 
The field $\varphi_{c(s)}$ gets ordered with the vacuum expectation values \cite{FO}
%%%%%%%%%%%%%%%%%%%%%%%%%%%%%%%%%%%%%%%%%%%%%%%%%%%%%%%%%%%%%%%%
\begin{equation}
\langle \varphi \rangle =\left\{ 
\begin{array}{l}
\sqrt{\displaystyle{\frac{\pi}{8K_{c(s)}}}}\hskip0.5cm(m_{c(s)} > 0) \\ 
0\hskip1.7cm(m_{c(s)} < 0)
\end{array}
\right. \,.  \label{orderedfields}
\end{equation}
%%%%%%%%%%%%%%%%%%%%%%%%%%%%%%%%%%%%%%%%%%%%%%%%%%%%%%%%%%%%%%%%
The ordering of these fields determines the symmetry properties of the
possible ordered groundstates of the fermionic system.

Using Eqs. (\ref{Kc})-(\ref{vcvs}) and (\ref{orderedfields})
one easily finds that there is a gap in the spin excitation spectrum 
$(M_{s}\rightarrow -\infty $) for $U+2W<0$. In the spin-gap sector the 
$\varphi _{s}$ field gets ordered with vacuum expectation value 
$\langle\varphi _{s}\rangle =0$. In the sector $U+2W\geq 0$ the spin 
excitations are gapless ($M_{s}\rightarrow 0$) and the low-energy properties 
of the spin sector are described by the free Bose field system with the 
fixed-point value of the parameter $K^{\ast}_{s}=1$. 

At half-filling the charge sector is gapless for $U,W \leq 0$ 
and along the line $U=2W$ for $U,W>0$.  At $U,W \leq 0$ the low-energy 
properties of the gapless charge sector are described by the free Bose field 
Hamiltonian (\ref{KGch}) with the fixed-point value of the parameter 
%%%%%%%%%%%%%%%%%%%%%%%%%%%%%%%%%%%%%%%%%%%%%%%%%%%%%
\begin{equation}
K^{\ast}_{c} \simeq 1 + \sqrt{2UW}/2\pi t.
\end{equation}
%%%%%%%%%%%%%%%%%%%%%%%%%%%%%%%%%%%%%%%%%%%%%%%%%%%%
For $U,W>0$ the line $U=2W$ ($m_{c}=0$) corresponding to the 
unstable fixed-point line $m_{c}=0, K_{c}-1<0$ (see Fig. \ref{fig:GSFD}). Here the infrared 
properties of the gapless charge sector are described by the free 
massless Bose field with the bare value of the Luttinger liquid parameter 
$K_{c}$. Moreover, the line $m_{c}=0, K_{c}-1<0$ ($U,W>0, U=2W$) separates two 
different insulating 
(charge gapped) sectors of the phase diagram: for $U > \max\{2W,0\}$, 
$m_{c}<0$, $M_{c}\rightarrow -\infty$ and therefore the 
$\varphi_{c}$ field gets ordered with vacuum expectation value 
$\langle\varphi_{c}\rangle =0$; for $2W > \max\{U,0\}$ $m_{c}>0$ 
$M_{c}\rightarrow +\infty$ and therefore the 
$\varphi_{c}$ field gets ordered with vacuum expectation value 
$\langle\varphi_{c}\rangle = \sqrt{\pi/8K_{c}}$. 
  
Away from half-filling the charge sector is gapless and is 
described by the free massless Bose field (\ref{KGch}). The corresponding 
correlations (\ref{freecorrelations1})-(\ref{freecorrelations2}) show a power 
law decay at large distances with critical indices determined by the bare 
value of the coupling constant $K_{c}$ (\ref{Kc}). The spin channel 
remains massive at $U+2W<0$ and gapless for $U+2W \geq 0$.

\subsection{\bf Correlation functions}

To clarify the symmetry properties of the ground states of the system in
different sectors we introduce the following set of order parameters
describing the short wavelength fluctuations of the site-located charge
and spin density
%%%%%%%%%%%%%%%%%%%%%%%%%%%%%%%%%%%%%%%%%%%%%%%%%%%%%%%%%%%%%%%%
\begin{eqnarray}\label{CDWop}
&\Delta_{{\small CDW}}(n)=
e^{i2\pi\nu n}\sum_{\sigma}\hat\rho_{\sigma}(n)\rightarrow & \nonumber\\
&\left\{ 
\begin{array}{l}
\sin \big(\sqrt{2\pi K_{c}}\varphi _{c}(x)\big)\cos \big(\sqrt{2\pi K_{s}}\varphi _{s}(x)\big), 
\text{ \ at \ \ }\nu =1/2 \\ 
e^{i\sqrt{2\pi K_{c}}\varphi _{c}(x)}\cos \big(\sqrt{2\pi K_{s}}\varphi _{s}(x)\big), \qquad 
\text{ \ at \ \ }\nu \neq 1/2
\end{array}
\right.&
\end{eqnarray}
%%%%%%%%%%%%%%%%%%%%%%%%%%%%%%%%%%%%%%%%%%%%%%%%%%%%%%%%%%%%%%%%
\begin{eqnarray}\label{SDWop}
&\Delta_{{\small SDW}}(n) = e^{i2\pi\nu n}\sum_{\sigma }\sigma\hat\rho_{\sigma}(n) 
\rightarrow &\nonumber\\
&\left\{ 
\begin{array}{l}
\cos \big(\sqrt{2\pi K_{c}}\varphi _{c}(x)\big)\sin \big(\sqrt{2\pi K_{s}}\varphi _{s}(x)\big),
\text{ \ at \ \ }\nu =1/2 \\ 
e^{i\sqrt{2\pi K_{c}}\varphi _{c}(x)}\sin \big(\sqrt{2\pi K_{s}}\varphi _{s}(x)\big),\text{
\ \ \ \ \ \ \ \ at \ \ }\nu \neq 1/2
\end{array}
\right.&
\end{eqnarray}
%%%%%%%%%%%%%%%%%%%%%%%%%%%%%%%%%%%%%%%%%%%%%%%%%%%%%%%%%%%%%%%%
and two superconducting order parameters corresponding to singlet ($\Delta
_{SS}$) and triplet ($\Delta _{TS}$) superconductivity: 
%%%%%%%%%%%%%%%%%%%%%%%%%%%%%%%%%%%%%%%%%%%%%%%%%%%%%%%%%%%%%%%%
\begin{eqnarray}
&&\Delta _{SS}(x) =R_{\uparrow }^{\dagger }(x)L_{\downarrow }^{\dagger}
(x)-R_{\downarrow }^{\dagger }(x)L_{\uparrow }^{\dagger }(x)\rightarrow \nonumber\\ 
&&\exp \left(i
\sqrt{\frac{2\pi }{K_{c}}}\theta _{c}(x)\right)\cos \big(\sqrt{2\pi K_{s}}\varphi _{s}(x)\big),
\label{SSop} \\
&&\Delta _{TS}(x) =R_{\uparrow }^{\dagger }(x)L_{\downarrow }^{\dagger
}(x)+R_{\downarrow }^{\dagger }(x)L_{\uparrow }^{\dagger }(x)\rightarrow \nonumber\\
&& \exp \left(i\sqrt{\frac{2\pi }{K_{c}}}\theta _{c}(x)\right) 
\sin\big(\sqrt{2\pi K_{s}}\varphi _{s}(x)\big).
\label{TSop}
\end{eqnarray}
%%%%%%%%%%%%%%%%%%%%%%%%%%%%%%%%%%%%%%%%%%%%%%%%%%%%%%%%%%%%%%%%
In the particular case of a half-filled band, we consider an additional pair
of order parameters, corresponding to the short wavelength fluctuations of
the bond-located charge and spin density
%%%%%%%%%%%%%%%%%%%%%%%%%%%%%%%%%%%%%%%%%%%%%%%%%%%%%%%%%%%%%%%%
\begin{eqnarray}
&&\Delta_{{\small Dimer}}(n) =(-1)^{n}\sum_{\sigma }(\hat{q}_{\sigma}(n) + 
\hat{q}^{\dagger}_{\sigma}(n))\rightarrow \nonumber \\
&&\cos \big(\sqrt{2\pi K_{c}}\varphi _{c}(x)\big)\cos \big(\sqrt{2\pi K_{s}}\varphi
_{s}(x)\big),  \label{DIMERop}
\end{eqnarray}
\begin{eqnarray}
&&\Delta_{Bd-SDW}(n) =(-1)^{n}\sum_{\sigma }\sigma (\hat{q}_{\sigma}(n) + 
\hat{q}^{\dagger}_{\sigma}(n))\rightarrow \nonumber \\
&&\sin \big(\sqrt{2\pi K_{c}}\varphi _{c}(x)\big)\sin \big(\sqrt{2\pi K_{s}}\varphi
_{s}(x)\big),  \label{Bd-SDWop}
\end{eqnarray}
%%%%%%%%%%%%%%%%%%%%%%%%%%%%%%%%%%%%%%%%%%%%%%%%%%%%%%%%%%%%%%%%

With these results for the excitation spectrum and the behavior of the corresponding fields 
Eqs. (\ref{freecorrelations1})-(\ref{orderedfields}) we now discuss the 
{\em weak-coupling} ground state phase diagram of the model (\ref{PKHmodel}). 

\subsection{\bf The half-filled band case.}

At half-filling the weak-coupling phase diagram of the 
PKH model consists of the following sectors (see Fig.2).  

\begin{itemize}
\item[A.] The Singlet Superconducting sector:\\
$U<0$ and $W<0$;\\
$M_{s}\neq 0$, $\langle \varphi _{s}\rangle =0$; \ 
$M_{c}=0$, $K^{\ast}_{c}>1$.
\end{itemize}

This sector of the coupling constants corresponds to the strong-coupling regime 
in the spin-channel and gapless charge channel. 
The dynamical generation of a spin gap accompanied by the ordering of the 
field $\varphi_{s}$ with vacuum expectation value 
$\langle \varphi_{s}\rangle =0$ leads to a complete suppression of the SDW, 
Bd-SDW, and TS instabilities. The CDW, Dimer, and SS instabilities survive 
and show a power-law decay at large distances
%%%%%%%%%%%%%%%%%%%%%%%%%%%%%%%%%%%%%%%%%%%%%%%%%%%%%
\begin{eqnarray}
\langle \Delta_{\small CDW}(x)\Delta_{\small CDW}(x^{\prime })\rangle &\sim& 
\langle \Delta_{\small Dimer}(x)\Delta_{\small Dimer}(x^{\prime })\rangle \nonumber\\
&\sim& \left| x-x^{\prime }\right|^{- K^{\ast}_{c}},
\end{eqnarray}
%%%%%%%%%%%%%%%%%%%%%%%%%%%%%%%%%%%%%%%%%%%%%%%%%%%%
%%%%%%%%%%%%%%%%%%%%%%%%%%%%%%%%%%%%%%%%%%%%%%%%%%%%%
\begin{equation}
\langle \Delta_{\small SS}(x)\Delta_{\small SS}(x^{\prime })\rangle 
\sim \left| x-x^{\prime }\right|^{-1/ K^{\ast}_{c}}.
\end{equation}
%%%%%%%%%%%%%%%%%%%%%%%%%%%%%%%%%%%%%%%%%%%%%%%%%%%%
where $K^{\ast}_{c}$ is the fixed-point value of the parameter $K_{c}$.
For $U, W < 0$,  $K^{\ast}_{c}>1$ and the SS instability dominates 
in the groundstate. 
%%%%%%%%%%%%%%%%%%%%%%%%%%%%%%%%%%%%%%%%%%%%%%%%%%%%%
\begin{figure}
\vspace{0mm}
\centerline{\psfig{file=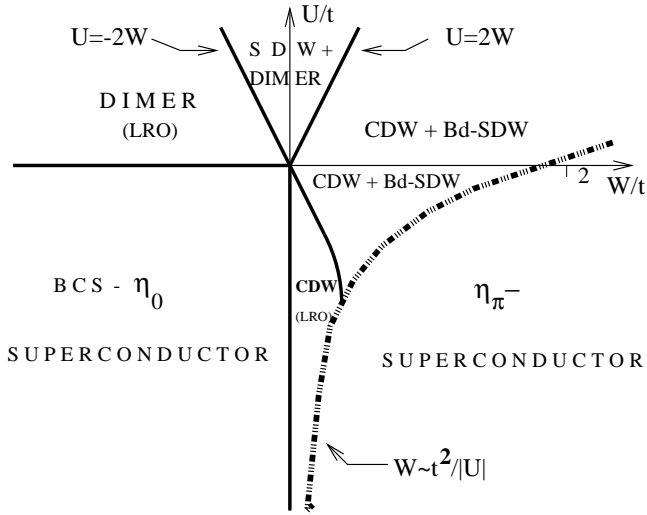,width=85mm,
silent=}}
\vspace{10mm}
\caption{The groundstate phase diagram of the 1D Penson-Kolb-Hubbard model 
at half-filling. The dot-dashed line marks a transition into the 
$\eta_{\pi}$-superconducting phase. Solid lines separate different phases: 
1.Dimer - long range ordered (LRO) dimerized (Peierls) phase. 2. 
CDW - LRO charge density wave phase; 
3. SDW $+$ Dimer - insulating phase with an identical 
power-law decay of spin-density-wave and Peierls correlations. 4. CDW $+$ 
Bd-SDW - insulating phase with an identical power-law decay of CDW and 
bond-located SDW correlations. 5. BCS - $\eta_{0}$-superconductor - singlet 
superconducting phase exhibiting a continuous evolution with increasing $|U|/t$ and/or 
$|W|/t$ 
from the standard BCS limit into the local pair $\eta_{0}$ state.}
\label{fig:gsPD1}
\end{figure}
%%%%%%%%%%%%%%%%%%%%%%%%%%%%%%%%%%%%%%%%%%%%%%%%%%

Note that, at $W=0$ or $U=0$, 
$K^{\ast}_{c}=1$ and the CDW, Dimer, and SS correlations show an identical 
power law decay at large distances. At $W=0$ coexistence of the CDW and superconducting instabilities 
in the ground state reflects the high $SU(2)\otimes SU(2)$ symmetry of the 1/2-filled 
Hubbard model \cite{FK}. In the case of the Penson-Kolb model ($U=0$), due to the 
charge-$U(1)$ symmetry of the pair-hopping term, there is no symmetry reason 
which guarantees scaling to the $SU(2)$ invariant fixed point 
$K^{\ast}_{c}=1$. As it was shown by Affleck and Marston, the higher order RG 
corrections give $K^{\ast}_{c}>1$ indicating the dominating character of the 
SS instability in the ground state of the 1/2-filled PK model for arbitrary 
$W<0$ \cite{AM}.

\begin{itemize}
\item[B.] The CDW sector:\\
$U<0$, $0<W<-U/2$.\\
$M_{s}\neq 0$, $\langle \varphi _{s}\rangle =0$; \ $M_{c}\neq 0$, 
$\langle \varphi_{c} \rangle =\sqrt{\displaystyle{\pi/8K_{c}}}$.
\end{itemize}

For $W>0$ the commensurability gap in the charge degrees of freedom opens. 
For $0<W<-U/2$ the spin sector remains 
massive. Ordering of the field $\varphi_{s}$ with vacuum expectation value 
$\langle \varphi_{s}\rangle =0$ leads to a complete suppression of the SDW, 
Bd-SDW, and TS instabilities. Ordering of the field $\varphi_{c}$
with vacuum expectation value 
$\langle \varphi_{c} \rangle =\sqrt{\displaystyle{\pi/8K_{c}}}$ 
leads to a suppression of the SS and Dimer correlations. The CDW 
correlations show a true long-range order 
%%%%%%%%%%%%%%%%%%%%%%%%%%%%%%%%%%%%%%%%%%%%%%%%%%%%%
\begin{equation}
\langle \Delta_{\small CDW}(x)\Delta_{\small CDW}(x^{\prime })\rangle \sim {\em 
const}
\end{equation}
%%%%%%%%%%%%%%%%%%%%%%%%%%%%%%%%%%%%%%%%%%%%%%%%%%%%
in the ground state. Therefore in this sector of the phase diagram the system 
displays the properties of the CDW insulator.

\begin{itemize}
\item[C.] The Peierls (dimerized) sector:\\
$U>0$, $0 < U < -2W$.\\
$M_{s}\neq 0$, $\langle \varphi _{s}\rangle =0$; \ $M_{c}\neq 0$, 
$\langle \varphi_{c} \rangle =0$.
\end{itemize}

Ordering of the field $\varphi_{s}$ with vacuum expectation value 
$\langle \varphi_{s}\rangle =0$ leads to a complete suppression of the SDW, 
Bd-SDW, and TS instabilities. Ordering of the field $\varphi_{c}$ with vacuum 
expectation value $\langle \varphi_{c} \rangle =0$ leads to a suppression of 
the SS and CDW correlations. The Dimer 
correlations show a true long-range order 
%%%%%%%%%%%%%%%%%%%%%%%%%%%%%%%%%%%%%%%%%%%%%%%%%%%%%
\begin{equation}
\langle \Delta_{\small Dimer}(x)\Delta_{\small Dimer}(x^{\prime })\rangle \sim {\em 
const}
\end{equation}
%%%%%%%%%%%%%%%%%%%%%%%%%%%%%%%%%%%%%%%%%%%%%%%%%%%%
in the ground state. Therefore in this sector of the phase diagram the system is a
dimerized (Peierls) insulator.

\begin{itemize}
\item[D.] The (CDW + Bd-SDW) sector:\\
$W>0$, $-2W < U < 2W$.\\
$M_{s}=0$, $K^{\ast}_{s}=1$, \ $M_{c}\neq 0$, 
$\langle \varphi_{c} \rangle =\sqrt{\displaystyle{\pi/8K_{c}}}$.
\end{itemize}

The generation of a gap in the charge excitation spectrum, 
accompanied by the ordering of the field $\varphi_{c}$ with vacuum 
expectation value 
$\langle \varphi_{c} \rangle = \sqrt{\displaystyle{\pi/8K_{c}}}$ 
leads to a suppression of the superconducting, SDW, and Dimer ordering. The CDW 
and Bd-SDW correlations show a power-law decay at large distances
%%%%%%%%%%%%%%%%%%%%%%%%%%%%%%%%%%%%%%%%%%%
\begin{eqnarray}  \label{correlCDW+Bd-SDW}
\langle \Delta_{{\small CDW}}(x) \Delta_{{\small CDW}}(x^{\prime})
\rangle &\sim & \langle\Delta_{{\small Bd-SDW}}(x)\Delta_{{\small Bd-SDW}
}(x^{\prime})\rangle  \nonumber \\
& \sim & \left| x - x^{\prime}\right|^{-1}.
\end{eqnarray}
%%%%%%%%%%%%%%%%%%%%%%%%%%%%%%%%%%%%%%%%%%
Therefore this sector of the phase diagram corresponds to the insulating phase 
with coexisting CDW and Bd-SDW instabilities.

\begin{itemize}
\item[E.] The (SDW + Dimer) sector:\\
$U>2|W|$.\\
$M_{s}=0$, $K^{\ast}_{s}=1$, \ $M_{c}\neq 0$, 
$\langle \varphi_{c} \rangle = 0$.
\end{itemize}

The generation of a gap in the charge excitation spectrum, accompanied by the 
ordering of the field $\varphi_{c}$ with vacuum expectation value 
$\langle \varphi_{c} \rangle = 0$ leads to a suppression of the superconducting, 
CDW, and Bd-SDW correlations. The SDW and Dimer correlations show a power-law 
decay at large distances
%%%%%%%%%%%%%%%%%%%%%%%%%%%%%%%%%%%%%%%%%%%
\begin{eqnarray}  \label{correlSDW+Dimer}
\langle \Delta_{{\small SDW}}(x) \Delta_{{\small SDW}}(x^{\prime})
\rangle &\sim & \langle\Delta_{{\small Dimer}}(x)\Delta_{{\small Dimer}
}(x^{\prime})\rangle  \nonumber \\
& \sim & \left| x - x^{\prime}\right|^{-1}.
\end{eqnarray}
%%%%%%%%%%%%%%%%%%%%%%%%%%%%%%%%%%%%%%%%%%
Therefore this sector of the phase diagram corresponds to the antiferromagnetic 
insulating phase with coexisting SDW and Peierls instabilities.

\subsubsection{\bf The non-half-filled band case.}
%%%%%%%%%%%%%%%%%%%%%%%%%%%%%%%%%%%%%%%%%%%%%%%%%%%%%
\begin{figure}
\vspace{0mm}
\centerline{\psfig{file=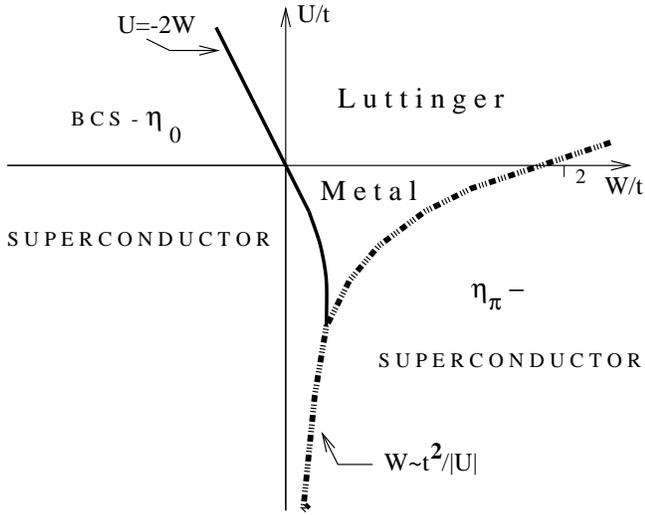,width=85mm,
silent=}}
\vspace{10mm}
\caption{The ground state phase diagram of the 1D Penson-Kolb-Hubbard model 
for $\nu \neq 1/2$. The dot-dashed line marks the transition into the 
$\eta_{\pi}$-superconducting phase. The solid line separates the superconducting phase 
from the Luttinger metal phase characterized by an identical 
power-law decay of the SDW, CDW, Bd-SDW, and Dimer correlations. }
\label{fig:gsPD2}
\end{figure}
%%%%%%%%%%%%%%%%%%%%%%%%%%%%%%%%%%%%%%%%%%%%%%%%%%
Away from half-filling and for $U<0$ the weak-coupling phase diagram of the 
PKH model consists of the following two sectors.  

\begin{itemize}
\item[A1.] The SS sector:\\
$U+2W<0$.\\  
$M_{s}\neq 0$, $\langle \varphi _{s}\rangle =0$;  \ $M_{c}= 0$, $K_{c}>1$.
\end{itemize}

In this sector of coupling constants the SDW, Bd-SDW, and TS instabilities 
are suppressed. The CDW, Dimer, and SS 
instabilities show a power-law decay at large distances
%%%%%%%%%%%%%%%%%%%%%%%%%%%%%%%%%%%%%%%%%%%%%%%%%%%%%
\begin{eqnarray}
\langle \Delta_{\small CDW}(x)\Delta_{\small CDW}(x^{\prime })\rangle &\sim& 
\langle \Delta_{\small Dimer}(x)\Delta_{\small Dimer}(x^{\prime })\rangle \nonumber\\
&\sim& \left| x-x^{\prime }\right|^{- K_{c}},
\end{eqnarray}
%%%%%%%%%%%%%%%%%%%%%%%%%%%%%%%%%%%%%%%%%%%%%%%%%%%%
%%%%%%%%%%%%%%%%%%%%%%%%%%%%%%%%%%%%%%%%%%%%%%%%%%%%%
\begin{equation}
\langle \Delta_{\small SS}(x)\Delta_{\small SS}(x^{\prime })\rangle 
\sim \left| x-x^{\prime }\right|^{-1/ K_{c}}.
\end{equation}
%%%%%%%%%%%%%%%%%%%%%%%%%%%%%%%%%%%%%%%%%%%%%%%%%%%%
As it follows from (\ref{Kc}), for $U=2W < 0$, $K_{c}>1$ and the SS 
instability dominates in the groundstate. 

\begin{itemize}
\item[B1.] The metallic Luttinger-liquid sector:\\
$U+2W \geq 0$.\\
$M_{s}= 0$, $K^{\ast}_{s}=1$, \ $M_{c}= 0$, $K_{c}<1$.
\end{itemize}

Both the charge and the spin excitations are gapless. All correlations show a 
power-law decay in the infrared limit. However, as far as $K_{c}<1$, the 
superconducting correlations
%%%%%%%%%%%%%%%%%%%%%%%%%%%%%%%%%%%%%%%%%%%
\begin{eqnarray}  \label{supercorrelLL}
\langle \Delta_{{\small SS}}(x) \Delta_{{\small SS}}(x^{\prime})
\rangle &\simeq & 
\langle\Delta_{{\small TS}}(x)\Delta_{{\small TS}}(x^{\prime})\rangle  
\nonumber \\
& \sim & \left| x - x^{\prime}\right|^{-1-1/K_{c}}
\end{eqnarray}
%%%%%%%%%%%%%%%%%%%%%%%%%%%%%%%%%%%%%%%%%%
decay faster then the density-density correlations 
%%%%%%%%%%%%%%%%%%%%%%%%%%%%%%%%%%%%%%%%%%%
\begin{equation}  \label{dencorrelLL}
\langle \Delta_{{\it i }}(x) \Delta_{{\it i}}(x^{\prime})\rangle
 \sim  \left| x - x^{\prime}\right|^{-1-K_{c}},
\end{equation}
%%%%%%%%%%%%%%%%%%%%%%%%%%%%%%%%%%%%%%%%%%%
where ${\it i} \equiv$ CDW,SDW,Dimer,Bd-SDW. Therefore in this sector the 
ground state of the PKH model shows properties of the Luttinger liquid phase 
with a weakly dominating tendency towards the density-density type ordering. 

To summarize this section,  we have presented the weak-coupling groundstate phase diagram 
for one-dimensional PKH model for arbitrary $U$ and $W$ 
($|U|,|W| \ll t$). We have shown that the model has a very rich phase diagram including at 
half-filling, the singlet-superconducting phase ($U, W<0$) and four different insulating 
phases corresponding to the Mott antiferromagnet ($U>2|W|$), the Peierls dimerized 
insulator ($0<U<-2W$) and the CDW insulator ($W>0, U<-2W$) and an unconventional insulating 
phase characterized by the coexistence of the CDW and the bond-located 
staggered magnetization Bd-SDW ($W>0, U<2|W|$). The possibility of bond-located ordering 
results from the site-off-diagonal nature of the pair-hopping term and is a special 
feature of the half-filled band case. Away from half-filling the phase diagram 
consists of the singlet-superconducting phase ($U+2W<0$) and the metallic Luttinger-liquid 
phase ($U+2W>0$).
 
Note the absence of the $\eta_{\pi}$-superconducting phase in the weak-coupling phase 
diagram. This results from the finite-band nature of the transition into an $\eta_{\pi}$-paired 
phase \cite{BJ}. Therefore in the used 
infinite-band bosonization approach, such a transition could not be traced. In the forthcoming section 
we will use exact Lanczos diagonalizations for chains up to $L=12$ sites to study in 
detail the nature of the transition into the $\eta_{\pi}$-superconducting phase. 

\section{Numerical studies}

In this section we study the nature of the transition into the $\eta_{\pi}$-superconducting 
phase using exact diagonalizations for chains up to $L=12$ sites. 

\subsection{The half-filled band case}

We start with the $W/t$ dependence of the groundstate energy for different $U \leq 0$. 
%%%%%%%%%%%%%%%%%%%%%%%%%%%%%%%%%%%%%%%%%%%%%%%%%%%%%
\begin{figure}
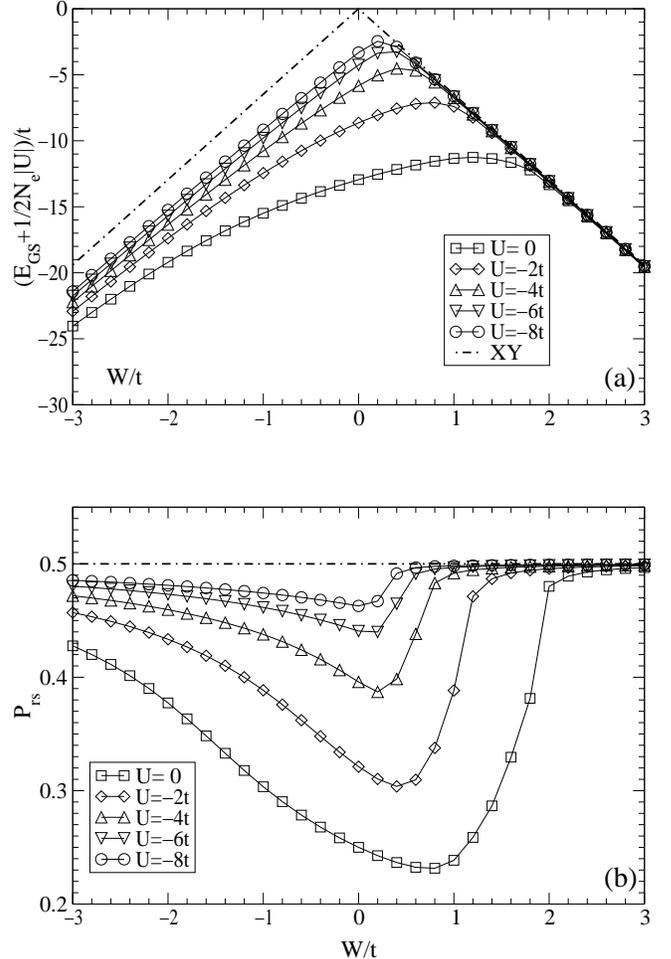

\vspace{0mm}
\centerline{\psfig{file=LPS1fig4a.eps,width=85mm,=}}
%silent=}}
\vspace{10mm}
\centerline{\psfig{file=LPS1fig4b.eps,width=85mm,=}}
%silent=}}
\vspace{10mm}
\caption{Groundstate energy $(E_{GS})$ shifted by $\frac{1}{2}N_{e}|U|$ \  (a) and real space pairing probability 
$P_{\em rs}$ (b) of the $L=10$ chain at half-filling, as a function of $W/t$, for 
$U/t=0,-2,-4,-6,-8$.  The corresponding symbols are indicated in the figure. The dot-dashed 
line in (a) represents the groundstate 
energy of the spin-$\frac{1}{2}$ $XY$ model corresponding to the limiting case 
$|W|/t$, $|U|/t\rightarrow\infty$.}
\label{fig:NR1}
\end{figure}
%%%%%%%%%%%%%%%%%%%%%%%%%%%%%%%%%%%%%%%%%%%%%%%%%%

In Fig. \ref{fig:NR1}a we have plotted the groundstate 
energy $(E_{GS})$ shifted by $\frac{1}{2}N_{e}|U|$ of the $L=10$ chain at half-filling as a function of $W/t$, for selected values of 
$U/t$. We clearly observe a different behavior of the 
groundstate energy for $W>0$ and $W<0$. For $W<0$ the ground state 
energy continuously approaches the XY model in the limit 
$\mid W\mid \rightarrow \infty $. There is no trace of any additional 
transition for $W<0$, in agreement with the density-matrix renormalization group 
(DMRG) results for $U=0$ \cite{SA}. However, for $W > 0$, the groundstate energy shows a non-monotonic behaviour. 
We observe that above some critical value of the pair-hopping coupling $W_{c}$ the energy is already 
almost linear in W and approaches the groundstate energy of the XY model. Hence for $W>W_{c}$ 
the {\em one particle hopping term is almost frozen out}. This change of behaviour is attributed to 
the transition to the $\eta_{\pi}$-superconducting state \cite{BJ}. The accurate definition for 
$W_{c}$ will be given below; however, the numerical data presented in Fig. \ref{fig:NR1}a already 
clearly indicate the renormalization of the critical value of the 
pair-hopping coupling $W_{c}$ by the on-site Hubbard attraction. At $U=0$ 
$W_{c} \simeq 1.8t$ \cite{BJ} and it reduces to $W_{c}\simeq 0.4t$ at $U/t=-8$. 

Another way to visualize this transition is to show the pairing phenomenon
in real space. The corresponding quantity is the expectation value to find an on-site pair and 
is given by 
%%%%%%%%%%%%%%%%%%%%%%%%%%%%%%%%%%%%%%%%%%%%%%%%%%%%%
\begin{equation}
P_{\em rs}=\frac{1}{L}\sum_{n=1}^{L} \langle \rho_{n\uparrow }\rho_{n\downarrow }\rangle.
\end{equation}
%%%%%%%%%%%%%%%%%%%%%%%%%%%%%%%%%%%%%%%%%%%%%%%%%%%%%
$P_{\em rs}$ is shown in Fig. \ref{fig:NR1}b. For $W<0$ the pairs appear 
continuously in the system. In the opposite case $W>0$, $W \ll t$ and $|U|<4t$ the 
tendency to pairing is even reduced. This ``effective repulsive'' character is absent for 
$U<-4t$, but remains up to $W<0.8t$ for $U=0$ and up to  $W<0.4t$ for $U=-2t$. We 
attribute this behaviour to a finite size effect where the energy scale coming from the 
exponentially small charge gap in $W/t$ is not traced in the $L=10$ chain. For 
$U<-4t$ the tendency to pairing is present for $W>0$. As is seen from Fig. \ref{fig:NR1}b 
for arbitrary $U<0$, 
the almost fully paired state is realized in the 
ground state of the half-filled PKH model for a {\it finite} value of the pair-hopping 
amplitude $W>W_{c}$. However, it should be stressed that due to the quasi-bosonic 
character of the pairs, the weight of unpaired particles remains {\it extremely small 
but finite} even for $W \gg W_{c}$.
The critical value of the pair-hopping amplitude corresponding to the transition into 
an $\eta_{\pi}$-superconducting state $W_{c}$ is straightforward to find when the number of 
electrons is $N_{e}=2(2n+1)$. The level crossing phenomenon is generic in this case. 
%%%%%%%%%%%%%%%%%%%%%%%%%%%%%%%%%%%%%%%%%%%%%%%%%%%%%
\begin{figure}
\vspace{0mm}
\centerline{\psfig{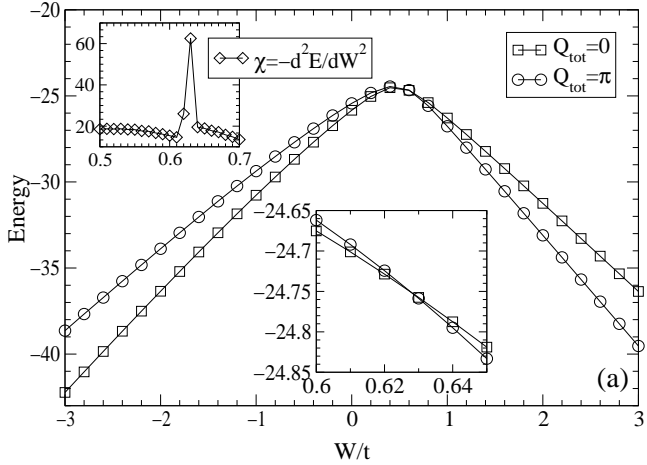}}
%silent=}}
\vspace{10mm}
\centerline{\psfig{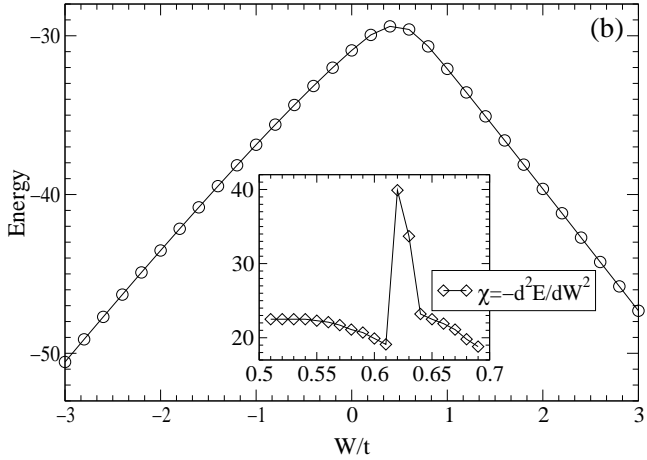}}
%silent=}}
\vspace{10mm}
\caption{(a) Lowest energy vs. $W/t$ calculated in the $Q_{tot}=0$ (open squares) and 
$Q_{tot}=\pi$ (open circles) subspaces in the case of a half-filled $L=10$ chain at 
$U/t=-4$. The insets in (a) show the level crossing and singularity in the generalized 
stiffness $\chi(W)=-\partial^{2}E_{0}(W)/\partial W^{2}$ at $W_{c} \simeq 0.625t$. (b) 
Groundstate energy vs. $W/t$ 
for a half-filled $L=12$ chain at $U/t=-4$. The inset in (b) shows the singularity in 
$\chi(W)$.}
\label{fig:NR2}
\end{figure}
%%%%%%%%%%%%%%%%%%%%%%%%%%%%%%%%%%%%%%%%%%%%%%%%%%
In  Fig. \ref{fig:NR2}a we have plotted the lowest energy levels (LEL) in each sectors $Q_{tot}=0$ and 
$Q_{tot}=\pi$ as a function of $W/t$ in the case of a half-filled $L=10$ chain and 
for $U/t=-4$. Indeed, in this case we observe a well defined transition from the $Q=0$ to the 
$Q=\pi$ sector at $W_{c}=0.625t$. At the same value we observe (see inset in Fig. \ref{fig:NR2}a) 
a singularity in the behaviour of the generalized stiffness (GS)
$$
\chi(W)=-\partial^{2}E_{0}/\partial W^{2}|_{U=const}
$$
due to the presence of a kink in the ground state energy $E_{0}(W)$ at the point where the LEL in 
sectors $Q_{tot}=0$ and $Q_{tot}=\pi$ cross each other. When the number of 
electrons is $N_{e}=4n$ the ground state of the system always remains in the sector of momentum 
space with $Q_{tot}=0$. In Fig. \ref{fig:NR2}b we have plotted the ground state energy of the 
half-filled PKH model for $L=12$. The insets show the anomaly in the behaviour of the generalized 
stiffness $\chi(W)$ at the same value $W_{c}=0.625t$ as in the case of $L=10$ chain. This clearly 
indicates the irrelevance of the finite-size effects already for $L=10$ in the considered case 
$U/t=-4$. We 
define the critical value of the pair-hopping amplitude corresponding to the transition 
into the $\eta_{\pi}$-paired state at the point where $\chi$ exhibits a singularity. 

%%%%%%%%%%%%%%%%%%%%%%%%%%%%%%%%%%%%%%%%%%%%%%%%%%%%%
\begin{figure}
\vspace{0mm}
\centerline{\psfig{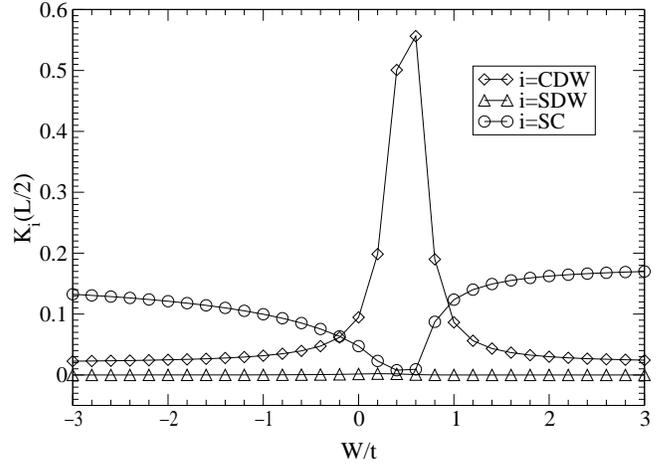}}
%silent=}}.
\vspace{10mm}
\caption{Correlation functions $K_{{\it i}}(r=L/2)$ ({\it i}= CDW, SDW, SC) vs. $W/t$ 
for the half-filled PKH chain ($L=10$) at $U=-4t$. The corresponding symbols are 
indicated in 
the figure. }
\label{fig:NR3}
\end{figure}
%%%%%%%%%%%%%%%%%%%%%%%%%%%%%%%%%%%%%%%%%%%%%%%%%%
To investigate the nature of ordering in the different phases we study the behaviour of the 
correlation functions. In Fig. \ref{fig:NR3} we plotted the absolute values of the spin-spin 
correlator 
%%%%%%%%%%%%%%%%%%%%%%%%%%%%%%%%%%%%%%%%%%%
\begin{equation}  \label{AFMcorr}
K_{SDW}(r) = \frac{1}{L} \sum_{n} \langle S^{z}(n)S^{z}(n+r) \rangle,
\end{equation}
%%%%%%%%%%%%%%%%%%%%%%%%%%%%%%%%%%%%%%%%%%%
the density-density correlator
%%%%%%%%%%%%%%%%%%%%%%%%%%%%%%%%%%%%%%%%%%%
\begin{equation}  \label{RhoRhocor}
K_{CDW}(r) = \frac{1}{L} \sum_{n} (-1)^{n}\langle \hat\rho (n)\hat\rho (n+r) \rangle,
\end{equation}
%%%%%%%%%%%%%%%%%%%%%%%%%%%%%%%%%%%%%%%%%%%
where $\hat\rho(n)=\sum_{\sigma}\hat\rho_{\sigma}$ and the superconducting correlator
%%%%%%%%%%%%%%%%%%%%%%%%%%%%%%%%%%%%%%%%%%%
%\begin{equation}  \label{SCcor}
%K_{SC}(r)  = \frac{1}{L} \sum_{n}\langle c_{n,\uparrow }^{\dagger}
%c_{n,\downarrow}^{\dagger }c_{n+r,\downarrow }c_{n+r,\uparrow } \rangle
%\end{equation}
%%%%%%%%%%%%%%%%%%%%%%%%%%%%%%%%%%%%%%%%%%%
%%%%%%%%%%%%%%%%%%%%%%%%%%%%%%%%%%%%%%%%%%%
\begin{equation}  \label{SCcor}
K_{SC}(r)  = \frac{1}{L} \sum_{n}\langle \eta^{\dagger}_{0}(n) 
\eta_{0}(n+r) \rangle
\end{equation}
%%%%%%%%%%%%%%%%%%%%%%%%%%%%%%%%%%%%%%%%%%%
at $r=L/2$ for different values of $W/t$ for the half-filled 
$L=10$ PKH chain at $U=-4t$. We clearly observe the existence of two 
superconducting phases separated by a phase with dominating density-density 
correlations. In contrast to generalized stiffness the finite-size effects in correlation 
functions are sufficiently strong for $L=10$ to use the large-distance properties of the 
correlation functions for the precise determination of the $W_{c}$. However, already in the case of 
the $L=10$ chain the essential physical details of the phase diagram are comletely transparent.

%%%%%%%%%%%%%%%%%%%%%%%%%%%%%%%%%%%%%%%%%%%%%%%%%%%%%
\begin{figure}
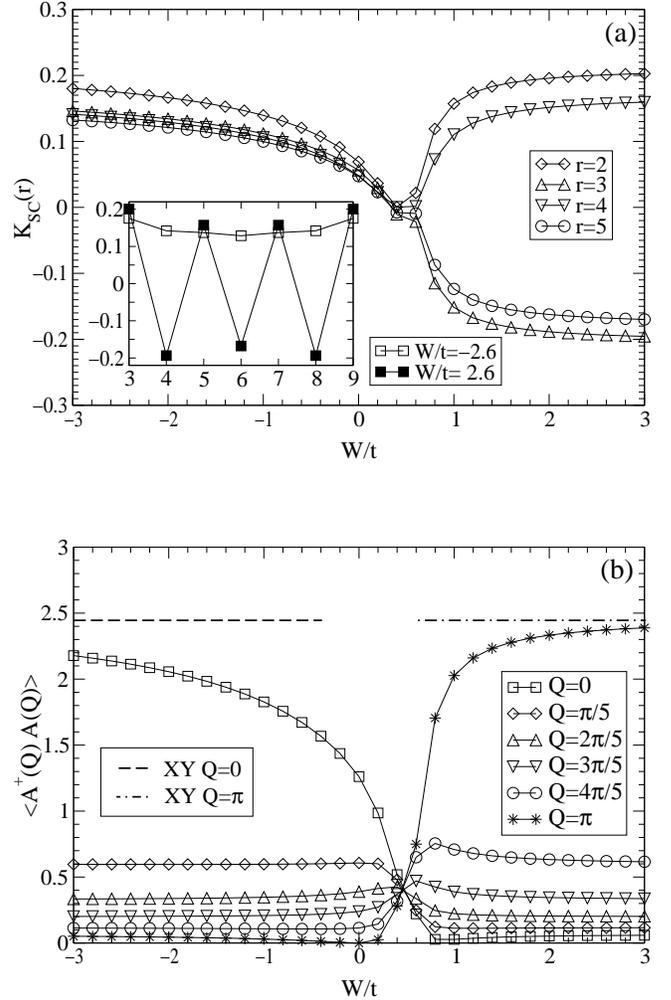

\vspace{0mm}
\centerline{\psfig{file=LPS1fig7a.eps,width=85mm,=}}
%silent=}}
\vspace{10mm}
\centerline{\psfig{file=LPS1fig7b.eps,width=85mm,=}}
%silent=}}
\vspace{3mm}
\caption{The superconducting correlator $K_{SC}(r)$ (a) and the distribution 
$P_{\em ms}(Q)=\langle A^{\dagger}_{Q}A_{Q} \rangle$ (b) vs. $W/t$ for the half-filled PKH chain 
($L=10$) at $U/t=-4$. The inset in (a) shows $K_{SC}(r)$ at $W/t=-2.6$ (open symbols) and $W/t=2.6$ 
(filled symbols). The dashed and dashed-dotted lines in (b) correspond to maximum values 
approached by the distribution $P_{\em ms}(Q)$ for $Q=0$ and $Q=\pi$ in the limiting case 
$|W|/t$, $|U|/t\rightarrow\infty$.}
\label{fig:NR4}
\end{figure}
%%%%%%%%%%%%%%%%%%%%%%%%%%%%%%%%%%%%%%%%%%%%%%%%%%
In order to clarify the nature of the superconducting phases corresponding to $W >0$ and $W <0$ 
respectively, we show in Fig. \ref{fig:NR4} the superconducting correlator $K_{SC}(r)$ (a) 
and the momentum space pairing probability 
$P_{\em ms}(Q) = \langle A^{\dagger}_{Q}A_{Q} \rangle$ (b) 
as a function of $W/t$, for $U/t=-4$ for the $L=10$ chain at half-filling.

For $W < 0$ $K_{SC}(r)$ smoothly increases with increasing $|W|/t$. The distribution $P_{\em ms}(Q)$ 
has a peak at $Q=0$. The weight of this peak continuously increases with 
$|W|/t \rightarrow \infty$. The probability to find a $\pi$-pair is almost zero. The inset in 
Fig. \ref{fig:NR4}a shows $K_{SC}(r)$ at 
$W/t=-2.6$ (open symbols). At $W/t=-2.6$ the superconducting correlations are well established and 
show weak decay with distance. In the opposite case, $W > 0$, $K_{SC}(r)$ and $P_{\em ms}(Q=0)$
continuously decrease. The superconducting correlations are completely suppressed at 
$0.4t<W<W_{c}\simeq 0.75t$. The expectation value to find a Cooper pair with center-of-mass momentum 
$Q=0$ becomes almost zero at $W>W_{c}$. However at $W>W_{c}$ we observe a drastic change of $K_{SC}(r)$ 
and $P_{\em ms}(Q)$. The superconducting correlations strongly increase and the strong peak at $Q= \pi$ 
appears spontaneously in the distribution $P_{\em ms}(Q)$. The probality to find a Cooper pair with 
center-of-mass momentum $Q=\pi$ quickly approaches its limiting value corresponding to the case 
$W/t\rightarrow\infty$. The inset in Fig. \ref{fig:NR4}a shows $K_{SC}(r)$ at $W/t=2.6$ (filled symbols). 
At $W/t=2.6$ the superconducting correlations are slightly stronger than at $W/t=-2.6$ and clearly 
show alternating behaviour with periodicity of two lattice spacing. The data presented in 
Fig. \ref{fig:NR4} clearly show the $\eta_{\pi}$ ordering at $W>W_{c}$.  

\subsection{The non-half-filled band case}

\begin{figure}
\vspace{0mm}
\centerline{\psfig{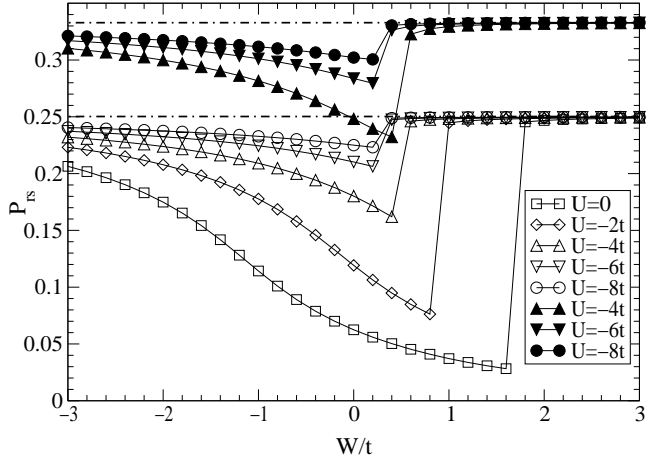}}
%silent=}}.
\vspace{10mm}
\caption{Real space pairing probability $P_{\em rs}$ vs. $W/t$ and various values of the parameter 
$U/t$ for $L=12$, $N_{e}=6$ (open symbols) and $N_{e}=8$ (filled symbols).}
\label{fig:NR5}
\end{figure}

In Fig. \ref{fig:NR5} we have plotted 
$P_{\em rs}$ as a function of $W/t$ for various values of the parameter $U/t$ calculated for 
the 
$L=12$ chain and two particular band fillings $\nu=1/4$ and $\nu=1/3$. As we see the 
transition into the $\eta_{\pi}$-paired state occurs also away from half-filling. Moreover, in 
contrast to the half-filled case, the transition is now very sharp. 
%%%%%%%%%%%%%%%%%%%%%%%%%%%%%%%%%%%%%%%%%%%%%%%%%%%%%
%%%%%%%%%%%%%%%%%%%%%%%%%%%%%%%%%%%%%%%%%%%%%%%%%%

\begin{figure}
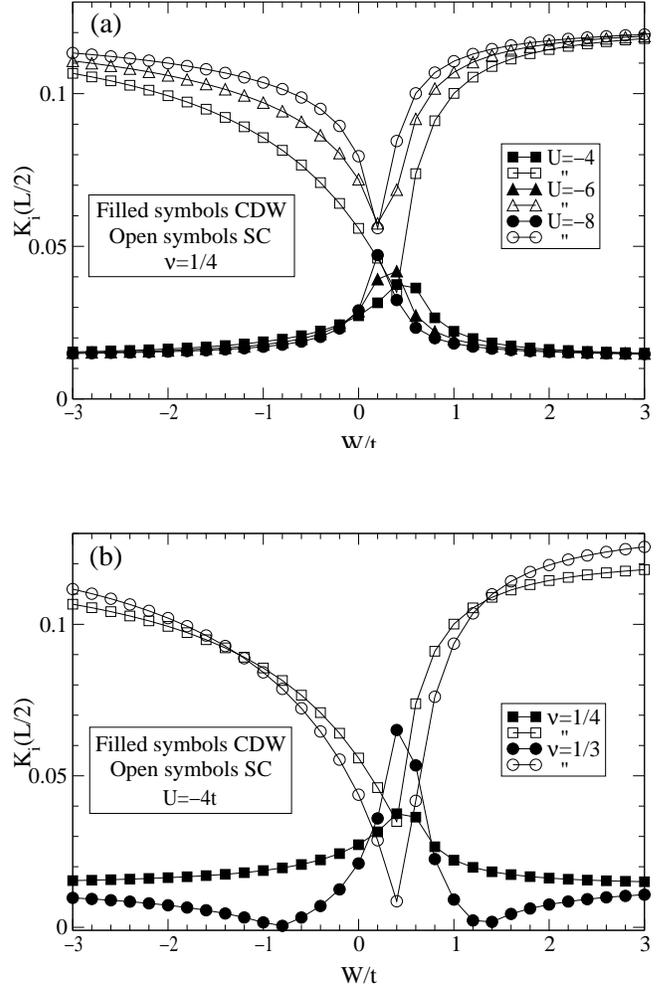

\vspace{0mm}
\centerline{\psfig{file=LPS1fig9a.eps,width=85mm,=}}
%silent=}}
\vspace{10mm}
\centerline{\psfig{file=LPS1fig9b.eps,width=85mm,=}}
\vspace{10mm}
\caption{Correlation functions $K_{CDW}(r=L/2)$ (filled symbols) and $K_{SC}(r=L/2)$ 
(open symbols) for the $L=12$ chain, $\nu=1/4$ and $U/t=-4,-6,-8$ (a) and for the $L=12$ chain, 
$U/t=-4$ and $\nu=1/4$ (squares) and $\nu=1/3$ (circles) (b).}
\label{fig:NR6}
\end{figure}

To study the phase diagram at $\nu \neq 1/2$ we investigate the behaviour of the CDW and SC 
correlations. In Fig. \ref{fig:NR6}a we have plotted the correlators $K_{CDW}$ and $K_{SC}$ at 
$r=L/2$ vs. $W/t$ calculated for the $L=12$ PKH chain in the case of $\nu=1/3$ and 
various values of the parameter $U/t$. As it follows from the behaviour of the correlation functions 
the sector with dominating density-density instability shrinks to a narrow strip with increasing 
on-site attraction. If for $U/t=-4$ the density-density correlations still dominate for 
$W \simeq 0.4t$, for $U/t=-6$ only superconducting correlations remain. 
%%%%%%%%%%%%%%%%%%%%%%%%%%%%%%%%%%%%%%%%%%%%%%%%%%%%%
%%%%%%%%%%%%%%%%%%%%%%%%%%%%%%%%%%%%%%%%%%%%%%%%%%

In Fig. \ref{fig:NR6}b we show the same correlators calculated for a fixed on-site attraction 
$U/t=-4$ and two different band fillings $\nu_{1}=1/4$ and $\nu_{2}=1/3$. As it is clearly seen 
the decreasing band filling leads to a reduction of the range of coupling 
constants where the system shows non-superconducting behaviour. In agreement with the 
strong-coupling expansion analysis of Sect. II we conclude that the system exhibits a transition 
from the BCS type singlet superconducting phase at $W<W_{c}$ into an $\eta_{\pi}$ paired 
state at $W>W_{c}$.

\subsection{Nature of the transition}
%%%%%%%%%%%%%%%%%%%%%%%%%%%%%%%%%%%%%%%%%%%%%%%%%%%%%
\begin{figure}
\vspace{0mm}
\centerline{\psfig{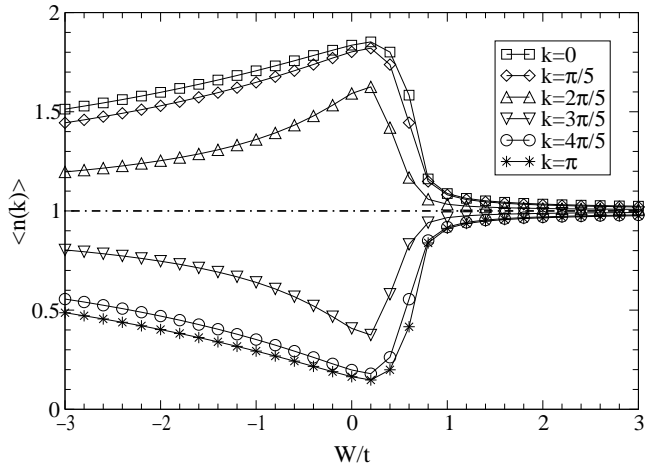}}
%silent=}}.
\vspace{10mm}
\caption{$\langle  n_{k} \rangle$ distribution for different values of the 
parameter $W/t$, in the case of half-filled $L=10$ chain and for $U/t=-4$. 
The corresponding symbols are indicated in the figure.}
\label{fig:NR7}
\end{figure}
%%%%%%%%%%%%%%%%%%%%%%%%%%%%%%%%%%%%%%%%%%%%%%%%%%

In order to show the finite bandwidth nature of the transition into an 
$\eta_{\pi}$-superconducting state in Fig. \ref{fig:NR7} we have plotted the momentum distribution 
%%%%%%%%%%%%%%%%%%%%%%%%%%%%%%%%%%%%%%%%%%%%%%%%%%%%%%%%%%%%%%%%%%%%%%%%%
$$
\langle n_{k}\rangle = \sum_{\sigma } 
\langle c_{k\sigma }^{\dagger }c^{\phantom{\dagger}}_{k\sigma }\rangle 
$$
%%%%%%%%%%%%%%%%%%%%%%%%%%%%%%%%%%%%%%%%%%%%%%%%%%%%%%%%%%%%%%%%%%%%%%%%%
in the ground state of the $L=10$ half-filled PKH chain as a function of the parameter $W/t$ at 
$U/t=-4$. In the case of an attractive Hubbard model 
($W=0, U=-4t$) the momentum distribution in the groundstate has a shape of the 
standard Fermi distribution: the states with $|k|<k_{F}=\pi/2$ are almost completely occupied, 
while the states with $|k|>k_{F}=\pi/2$ are almost empty. In the case of $W<0$ the 
momentum distribution remains qualitatively unchanged. With increasing $|W|/t$ the distribution 
$\langle n_{k}\rangle$ in a monotonic way approaches its limiting behaviour 
$\langle n_{k}\rangle =1$  at $|W|/t\rightarrow\infty$. 

At $W>0$ the standard shape in the momentum distribution remains for 
$0<W<W_{c}$; however, at $W \simeq W_{c} \simeq 0.625t$ we observe an abrupt change in the momentum 
distribution. For $W > W_{c}$ $\langle  n_{k} \rangle  \simeq 1$ for all $k$. States with different 
momenta are almost equally occupied by electrons. There is no trace of the Fermi distribution. As far 
as states 
with momentum $k$ and $\pi -k$ are almost equally occupied after the transition, the contribution of 
the {\em one-electron band} to the groundstate energy is almost {\em completely suppressed}. The 
ground state energy of the system becomes linear in $W$. Therefore we conclude that the transition 
into an $\eta_{\pi}$-superconducting state corresponds to {\em destruction of the single-electron 
conduction band} and the {\em creation of a strongly-correlated $\eta_{\pi}$-pair band}. 

%%%%%%%%%%%%%%%%%%%%%%%%%%%%%%%%%%%%%%%%%%%%%%%%%%%%%
\begin{figure}
\vspace{0mm}
\centerline{\psfig{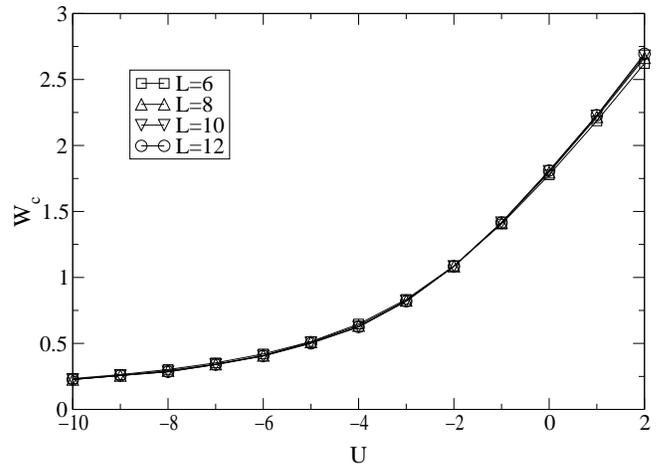}}
%silent=}}.
\vspace{10mm}
\caption{$W_{c}$ vs. $U$ calculated for the half-filled PKH chain with $L=6,8,10,12$.}
\label{fig:NR8}
\end{figure}
%%%%%%%%%%%%%%%%%%%%%%%%%%%%%%%%%%%%%%%%%%%%%%%%%%

As far as the transition into an $\eta_{\pi}$-paired state is connected with the formation of 
a $\eta_{\pi}$-pair (doublon) band, after the transition the characteristic length scale in the 
system $\xi \simeq 0$ . Therefore the finite-size effects are very weak. In Fig. \ref{fig:NR8} we have 
plotted $W_{c}$ vs. $U$ calculated for the half-filled chain with $L=6,8,10,12$.  As it can be seen 
in this figure the finite size effects have only a {\it very weak} influence on the critical value. 

\section{Summary}

To summarize, in this paper we have studied the groundstate phase diagram of 
the one-dimensional Penson-Kolb-Hubbard model using the continuum-limit field theory
approach and finite system numerical studies. We have shown that the model has a very rich 
phase diagram including, at half-filling (see Fig.\ref{fig:gsPD1}), two different superconducting 
phases and four different insulating phases corresponding to the singlet superconducting 
phase ($U,W<0$) with BCS type pairing at $|U|,|W| \ll t$, monotonically evolving into 
the local-pair $\eta_{0}$-superconducting phase with $U/t$ and/or $W/t$ 
approaching $-\infty$; the $\eta_{\pi}$-superconducting phase ($W>W_{c}(U)>0$); the Mott 
antiferromagnet ($U>2|W|$) phase, the Peierls dimerized insulator ($0<U<-2W$); the 
CDW insulator ($W>0, U<-2W$); the unconventional insulating phase characterized with 
coexistence of the CDW and bond-located staggered magnetization Bd-SDW 
($W >0, U<2|W|$). 

Away from half-filling (see Fig.\ref{fig:gsPD2}) the phase diagram consists of the 
singlet-superconducting phase ($W < \min\{-U/2,W_{c}(U)\}$), and the metallic 
Luttinger-liquid (LL) phase ($-U/2< W < W_{c}(U)$), and the $\eta_{\pi}$-superconducting phase 
($W > W_{c}(U)>0$). With increasing on-site attraction the LL phase shrinks and at a critical 
value $U_{c}$, determined by the condition $U_{c}+2W_{c}(U_{c})=0$, only the critical line 
$W_{c}(U)$ remains for $U<U_c$. In this range of couplings the transition from an 
$\eta_{0}$-superconductor to an $\eta_{\pi}$-superconductor is realized. 
The critical value $W_{c}(U)$ weakly depends on the band filling displaying a non-monotonic 
behaviour: at $|U|/t \ll 1$ a slight decrease while at $|U|/t \gg 1$ a slight increase 
with increasing $\nu$.

The obtained phase diagram is in agreement with results of previous 
studies based on the real space renormalization group method \cite{BR}, the continuum-limit 
approach \cite{JM}, the slave-boson mean-field method \cite{Rob} as well as for the 
particular case of $U=0$ with the results of numerical studies using exact (Lanczos)
diagonalizations \cite{BJ} and DMRG studies 
\cite{SA}.

Using Lanczos diagonalization we have studied in detail the transition into the 
$\eta_{\pi}$-paired state. After the transition the on-site pairing probability sharply 
approaches its limiting value corresponding to the case $W/t\rightarrow\infty$ and the 
$\eta_{\pi}$-superconducting correlations dominate in the system. The transition corresponds to an 
abrupt change in the groundstate structure. After the transition the single-electron 
conduction band is completely destroyed and the strongly-correlated $\eta_{\pi}$-pair (doublon) 
band is established. The transition occurs at any band filling, the critical value $W_{c}(U)$ {\em 
weakly depends on the band filling} but is strongly renormalized by the on-site Hubbard 
interaction: $W_{c}(U)$ is linear in $U$ for $|U|/t$ of the order of unity and is inversely 
proportional to $|U|$ i.e $W_{c}(U) \simeq -t^{2}/|U|$ for $|U|/t \gg 1$. We believe that this 
phase diagram is generic for the Penson-Kolb-Hubbard model and will remain unchanged in higher 
dimensions. 

This work was partially supported by the DFG through SP 1073. G.J. was 
partially supported by the INTAS-Georgia grant N 97-1340. M.S acknowledges 
the hospitality at the Institute of Physics, University of Augsburg, where 
part of this work was performed.

\end{document}